\documentclass[12pt,preprint]{aastex}
\def\ltsima{$\;\buildrel < \over \sim \;$}
\def\simlt{\lower.5ex \hbox{\ltsima}}
\def\gtsima{$\;\buildrel > \over \sim \;$}
\def\simgt{\lower.5ex \hbox{\gtsima}}

\shorttitle{Interstellar fluorine chemistry}
\shortauthors{Neufeld, Wolfire \& Schilke}

\begin{document}

\title{The chemistry of fluorine-bearing molecules in diffuse and dense interstellar gas clouds}
\author{David A. Neufeld\altaffilmark{1}, Mark G. Wolfire\altaffilmark{2}, and Peter Schilke\altaffilmark{3}}
\altaffiltext{1}{Department of Physics and Astronomy, Johns Hopkins University,
3400 North Charles Street, Baltimore, MD 21218; neufeld@pha.jhu.edu}
\altaffiltext{2}{Department of Astronomy, University of Maryland, College Park, MD 20742; mwolfire@astro.umd.edu}
\altaffiltext{3}{Max-Planck-Institut f\"ur Radioastronomie, Auf dem H\"ugel 69, 53121 Bonn, Germany; schilke@mpifr-bonn.mpg.de}

\begin{abstract}

We present a theoretical investigation of the chemistry of fluorine-bearing molecules in diffuse and dense interstellar gas clouds, combining recent estimates for the rates of relevant chemical reactions with a self-consistent model for the physical and chemical conditions within gas clouds that are exposed to the interstellar ultraviolet radiation field.  The chemistry of interstellar fluorine is qualitatively different from that of any other element, because -- unlike the 
neutral atoms of any other element found in diffuse or dense molecular clouds -- atomic fluorine undergoes an exothermic reaction with molecular hydrogen.  Over a wide range of conditions attained within interstellar gas clouds, the product of that reaction -- hydrogen fluoride -- is predicted to be the dominant gas-phase reservoir of interstellar fluorine nuclei.  
Fluorine is the heavy element which shows the greatest tendency toward molecule formation; in diffuse clouds of small extinction, the predicted HF abundance can even exceed that of CO, even though the gas-phase fluorine abundance is four orders of magnitude smaller than that of carbon.  Our model predicts HF column densities $\sim 10^{13} \,\rm cm^{-2}$ in dark clouds and column densities
as large as $10^{11} \,\rm cm^{-2}$ in diffuse interstellar gas clouds with total visual extinctions as small as 0.1 mag.  Such diffuse clouds will be detectable by means of absorption line spectroscopy of the 
$J=1-0$ transition at 243.2~$\mu$m using the Stratospheric Observatory for Infrared Astronomy (SOFIA) and the Herschel Space Observatory (HSO).  
The CF$^+$ ion is predicted to be the second most abundant fluorine-bearing molecule, with typical column densities a factor $\sim 10^2$ below those of HF; with its lowest two rotational transitions in the millimeter-wave spectral region, CF$^+$ may be detectable from ground-based observatories.  
HF absorption in quasar spectra is a potential probe of molecular gas at high redshift, providing a possible bridge between the UV/optical observations capable of probing H$_2$ in low column density systems and the radio/millimeter-wavelength observations that probe intervening molecular clouds of high extinction and large molecular fraction; at redshifts beyond $\sim 0.3$, HF is potentially detectable from ground-based submillimeter observatories in several atmospheric transmission windows.

\end{abstract}


\section{Introduction}

To date, molecules containing the ten elements hydrogen, carbon, nitrogen, oxygen, fluorine, silicon, phosphorus, sulfur, chlorine, and iron have been detected in the interstellar medium (ISM).\footnote{In addition, the molecules containing the elements Na, Mg, Al, and K have been detected in circumstellar outflows.}  While the chemical processes leading to the formation and destruction of molecules containing the more abundant of these elements (e.g. H, C, N, O, Si, S) have been the subject of extensive theoretical investigation over a period of several decades, the chemistries of the less abundant elements have received far less attention.  In particular, no systematic theoretical study of the chemistry of interstellar fluorine-bearing molecules has yet been undertaken, although a preliminary investigation undertaken following the discovery of interstellar hydrogen fluoride (Neufeld et al.\ 1997; hereafter NZSP) suggested a strong tendency towards the formation of HF; this investigation suggested that HF would be the overwhelmingly dominant reservoir of gas-phase F nuclei over a wide range of physical conditions, and led NZSP to argue that observations of HF provided a particularly useful probe of the depletion of interstellar molecules onto icy grain mantles within cold, shielded regions of the ISM.

In this paper, we report the first systematic study of the chemistry of interstellar fluorine.  This study is motivated by several recent and current developments: (1) the discovery of interstellar hydrogen fluoride in 1997 (NZPS); (2) new quantal calculations of the rate coefficient for $\rm F + H_2 \rightarrow H + HF$ (Zhu et al. 2002), which strengthen further the case for large HF abundances over a wide range of interstellar conditions; (3) the prospects for the widespread detection of HF absorption with upcoming airborne and satellite submillimeter telescopes; (4) the potential utility of HF observations as a probe of the depletion of interstellar molecules onto icy grain mantles.  

In \S2, we present a discussion of the fundamental chemical and photochemical processes leading to the formation and destruction of fluorine-bearing molecules in interstellar gas clouds.  In \S3, we discuss a model for the chemistry of fluorine in diffuse and dark clouds irradiated from the outside by ultraviolet radiation.  Predicted abundances and column densities of fluorine-bearing species are presented for a range of physical parameters relevant to the interstellar medium.  A discussion of these results follows in \S4, with particular emphasis on the prospects for future observations.

\section{Interstellar fluorine chemistry}

\subsection{Overview: thermochemistry of F-bearing species}

The large dissociation energy of hydrogen fluoride makes the chemistry of interstellar fluorine qualitatively different from that of any other element.
This key feature of the fluorine chemistry is made clear in Table 1, which lists the standard heats of formation of the hydrides and fluorides of the elements C, N, O, Si and S.  The values presented here are relative to the enthalpies of formation of the constituent atoms; the tabulated quantity is therefore $\Delta_f H^0_{\rm gas} ({\rm XY}) - \Delta_f H^0_{\rm gas} ({\rm X}) - \Delta_f H^0_{\rm gas} ({\rm Y})$.  In each case, the enthalpies recommended by the NIST Chemistry Web book\footnote{NIST Standard Reference Database Number 69, Eds. P.~J. Linstrom and W.~G. Mallard, March 2003, National Institute of Standards and Technology, Gaithersburg MD, 20899; available on-line at http://webbook.nist.gov} were adopted.  The values presented for molecular ions make use of the ionization potentials recommended by the NIST Chemistry Web book, and any differences between the specific heat capacity of each ion and its corresponding neutral were neglected.

The thermochemical data presented in Table 1 show that atomic fluorine -- unlike C, C$^+$, N, O, Si, Si$^+$, S or S$^+$ -- can undergo exothermic reaction with H$_2$ to form a diatomic hydride.  Thus, the  reaction of the dominant ionization state of fluorine (F) with the dominant molecular constituent of the interstellar gas (H$_2$) potentially provides a uniquely efficient pathway to the formation of HF. 
The chemistry of interstellar chlorine shows some similarity to that of fluorine, in 
that $\rm Cl^+$, the dominant ionization state of chlorine in atomic clouds, reacts exothermically with H$_2$ to form $\rm HCl^+$.  A critical difference between the two cases, however, is that $\rm HCl^+$ -- unlike HF of course -- undergoes dissociative recombination to produce Cl.  Thus atomic chlorine, which does not react exothermically with H$_2$, is predicted to become the dominant ionization stage at precisely the point at which the molecular fraction becomes large.

A second implication of the data presented in Table 1 is that HF -- once formed -- is not easily destroyed.  Reactions with C, N, O, 
Si, S and S$^+$ are all substantially endothermic and therefore negligibly slow at the low temperatures that characterize molecular clouds.  Table 2 presents the heats of reaction for 20 possible reactions that might destroy HF.  Of these, only reactions with C$^+$, 
CH, CH$^+$, Si$^+$, SiH, and SiH$^+$ are exothermic\footnote{In NZSP, we erroneously included Si$^+$ in the list of species that cannot react exothermically with HF}.  Given the typical temperatures and chemical composition in molecular clouds, the reactions of HF with C$^+$ and Si$^+$ are expected to be the most important of the those listed in Table 2, leading to the stable molecular ions CF$^+$ and SiF$^+$ that are isoelectronic with CO.

In Table 3, proton affinities are compared for several astrophysically important molecules.  Here again, rather few proton transfer reactions that could destroy HF can proceed exothermically: at the low temperatures in molecular clouds, only H$_3^+$ and H$_2^+$ can react with HF to form H$_2$F$^+$.  Furthermore, the ionization energy of HF (16.03~eV) being larger than that of H, HF does not undergo exothermic charge transfer with H$^+$ 
and -- like atomic fluorine -- cannot be photoionized by radiation longward of the Lyman limit.

In the remainder of this section, we present a detailed discussion of the primary formation and destruction mechanisms for hydrogen fluoride.

\subsection{Formation of HF}

The reaction 
$$\rm F + H_2 \rightarrow HF + H \eqno({\rm R1}) $$
has been the subject of intensive experimental and observational study, the results of which have been reviewed recently -- and added to -- by Zhu et al.\ (2002).  The reaction has an activation energy barrier $\sim 4\, \rm kJ \, mol^{-1}$, although the reaction rate is significantly enhanced at low temperatures by tunneling and is finite at zero temperature (Balakrishnan \& Dalgarno 2001).  Thus, at the temperatures $T \sim 10 - 100$~K typical of molecular clouds, the rate coefficient was substantially underestimated by the simple Arrenhius  expression that we adopted in NZSP to extrapolate from laboratory measurements obtained at higher temperatures.  Here instead, we adopt the results of the detailed calculations obtained by Zhu et al. (2002) and kindly made available to us for temperatures (10 -- 50~K) outside the range computed previously.   
At the densities typical of the interstellar medium, atomic fluorine is found primarily in the ground $^2P_{3/2}$ fine-structure state (Zhu et al.\ 2002), which has a higher rate coefficient 
than $^2P_{1/2}$ for reaction with H$_2$.  The overall rate of reaction of H$_2$ with F is therefore well-approximated by the specific rate for reaction with F\,($^2P_{3/2}$).  Figure 1 shows the rate coefficients computed by Zhu et al.\ for the reaction of F\,($^2P_{3/2}$) with H$_2$ $J=0$ (red crosses) and $J=1$ (blue crosses). 
Black crosses show a weighted average applicable when the H$_2$ populations are in local thermodynamic equilibrium (LTE)\footnote{The exact ortho-to-para ratio for H$_2$ has been the subject of considerable theoretical and observational study.  In the diffuse molecular clouds, the observed ratio ${\rm H}_2\,\,J=1 \, / \,{\rm H}_2\,\,J=0$ is generally consistent with LTE (e.g. Savage et al.\ 1977), although the value in dark clouds is less well constrained.  Fortunately, in this case, the reaction rates for H$_2\,\, J=1$ and H$_2\,\,J=0$ are not that dissimilar.}
, and the solid black curve shows our adopted fit to that average (see Table 4), which is accurate to better than $5\%$ throughout the temperature range plotted.  

Some remaining uncertainty is associated with the F + H$_2$ potential energy surface (PES).  As Zhu et al.\ and others have noted, a more recent PES that has been computed with account taken of the spin-orbit interaction (Castillo et al.\ 1998; hereafter the HSW potential) appears to yield a {\it poorer} fit to the experimental data than an earlier PES for which the spin-orbit interaction was neglected (Stark \& Werner 1996; hereafter SW).  In particular, the barrier in the HSW potential appears to be too high.  Accordingly, Zhu et al.\ adopted the earlier SW PES in the calculations that they presented.

In NZSP, we argued that the reaction of F with H$_2$O might provide an important additional HF formation route at low temperatures ($T \le 30$~K) for which the F + H$_2$ reaction rate was small.  The rate coefficient for the reaction
$$\rm F + H_2O \rightarrow OH + HF  \eqno({\rm R2})$$
has been measured by Stevens et al.\ (1989) as $\rm  (1.6 \pm 0.3) \times 10^{-11}
cm^3 s^{-1}$.  The rate coefficient showed no detectable variation over the temperature range 240--373~K and the implied limit on any activation energy barrier was ($-28 \pm 42$)~K. 
Here, we adopt the same rate coefficient.  The importance of reaction (R2), however, is diminished because the adopted rate coefficient for (R1) is now much larger than what we previously assumed in the low temperature regime.

\subsection{Destruction of HF}

\subsubsection{Ion-neutral reactions}

As discussed in \S 2.1 above, HF can react exothermically with the ions C$^+$, Si$^+$ and H$_3^+$ in the reactions
$$\rm C^+ + HF \rightarrow CF^+ + H  \eqno({\rm R3})$$
$$\rm Si^+ + HF \rightarrow SiF^+ + H  \eqno({\rm R4})$$
$$\rm H_3^+ + HF \rightarrow H_2 + H_2F^+  \eqno({\rm R5})$$
In addition, dissociative charge transfer with He$^+$ is also exothermic:
$$\rm He^+ + HF \rightarrow H + F^+ + He \eqno({\rm R6}) $$
For molecules with large dipole moments such as HF, the rates of ion-neutral reactions are expected (e.g.\ Adams, Smith \& Clary, 1985) -- and indeed observed (e.g.\ Rowe et al.\ 1985) -- to increase significantly with decreasing temperature.  To our knowledge, none of the reactions (R3)-- (R6) has been measured at interstellar temperatures.  In estimating the rate coefficients for these reactions, we have made use of the statistical adiabatic capture model (SACM; Troe 1985, 1987, 1996) to determine the rate of ion-dipole capture, together with the 
assumption that the reaction probability following capture is unity.
In this approximation, the capture rate depends solely upon the dipole moment, $\mu_D= \rm 1.826\,Debye $; the polarizability, $\alpha = 2.4 \times 10^{-24} \rm \, cm^3$; the rotational constant, $B = 20.54 \rm \, cm^{-1}$ ; and the reduced mass of the reacting system, $\mu$.

The capture rate is given in the limit of low temperature ($T \simlt hcB/k$), by the expression 
$$k_{a} = k_L (1 + \mu_D^2/3\alpha hcB)^{1/2},\eqno(1)$$ 
where $k_L = 2 \pi \vert q \vert (\alpha/\mu)^{1/2}$ $=3.6 \times 10^{-9} \rm cm^3 s^{-1}$ is the Langevin rate coefficient for reactions involving an ion of charge $q$ (i.e.\ the value that would obtain in the absence of any permanent dipole moment.)  
Gaussian electrostatic units are adopted here, 
with $\alpha$ having dimensions of length$^3$ and $\mu_D$ having units of e.s.u.~$\times$~length.  At temperatures $T \simgt hcB/k$, the reaction rate can approximated by 
$$k_{b}= k_L (1 + f \mu_D (2/\pi \alpha k T)^{1/2}) \eqno(2)$$  
(Troe 1987; using the simplified analytic expressions in the Appendix),
where $f \sim 0.4$ if the HF rotational states are thermally-populated and $f \sim 1$ if HF is rotationally cold.  (This high-temperature case is often referred to as the average dipole orientation (ADO) approximation and the specific case $f=1$ as the locked-dipole approximation.)  The low and high temperature regimes are bridged by the expression 
$$k_{cap} = k_b \,[1 - {\rm exp}\, (-k_a/k_b)] \eqno(3)$$

Using this model, we computed the rate cofficients for reactions (R3) -- (R6) over the temperature range 10 to 100~K and then fit a power-law to the temperature dependence. Because interstellar densities are typically much smaller than the critical density\footnote{The critical density at which the rate of collisional de-excitation equals the spontaneous radiative decay rate
is $n(\rm {H}_2) \sim 10^{10}\, {\rm cm}^{-3}$ at 50~K, given the collisional rate coefficients computed by Reese et al.\ (2005)} at which the HF $J=1$ rotational state reaches its LTE population, we considered the case where HF is rotationally-cold (i.e.\ entirely in the ground state, $J=0$.)  The result of the power-law fit was
$$k_{cap} = 2.0 \times 10^{-8} (T/300{\rm K})^{-0.15} (\mu/{\rm a.m.u.})^{-1/2} \rm cm^3 s^{-1}\eqno(4)$$
and maximum error in the power-law fit was $12\%$ over the temperature range 10 -- 100 K.

Fluorine ions produced in reaction (R6) can undergo a series of two exothermic hydrogen atom abstraction reactions to form H$_2$F$^+$:
$$ \rm F^+ + H_2 \rightarrow HF^+ + H \eqno({\rm R7})$$
$$ \rm HF^+ + H_2 \rightarrow H_2F^+ + H \eqno({\rm R8})$$
We adopt the estimates (see Table 4) provided by Le Teuff, Millar \& Markwick (2000; a.k.a.\ the UMIST ratefile) for the rate coefficients of the analogous reactions involving the element chlorine.  The exact value of these rate coefficients has no bearing upon the overall HF abundance because every F$^+$ ion produced by reaction (R6) is rapidly converted to $\rm H_2F^+$.
Reaction of CF$^+$ with H$_2$ to form HCF$^+$ is endothermic by 180~kJ~mol$^{-1}$ and is therefore negligibly slow at the temperatures of interest.    
The molecular ions produced by reactions (R3) -- (R8) are destroyed by dissociative recombination (DR) via the reactions
$$\rm CF^+ + e \rightarrow C + F \eqno({\rm R9})$$
$$\rm SiF^+ + e \rightarrow Si + F \eqno({\rm R10})$$
$$\rm HF^+ + e \rightarrow H + F \eqno({\rm R11})$$
$$\rm H_2F^+ + e \rightarrow HF + H \eqno({\rm R12})$$
$$\rm \phantom{H_2F^+ + e} \rightarrow F + 2H$$
$$\rm \phantom{H_2F^+ + e} \rightarrow F + H_2$$ 

We are unaware of any experimental or theoretical studies that specifically address the rates of these reactions.  We therefore adopt a rate coefficient typical of other DR reactions: $2 \times 10^{-7}\,(T/300)^{-1/2}\,\rm cm^3\,s^{-1}$ for diatomic ions [(R9) -- R(11)] and $7 \times 10^{-7}\,(T/300)^{-1/2}\,\rm cm^3\,s^{-1}$ for polyatomic ions [reaction (R12)].  The exact value adopted does not affect the predicted HF abundance (although, of course 
the minor species $\rm CF^+$, $\rm SiF^+$, and $\rm H_2F^+$ will have abundances that vary inversely with their assumed DR rate coefficients).  The branching ratio for the DR of $\rm H_2F^+$ will affect the effective rate of HF destruction via reactions (R5) and (R6), because HF is immediately reformed unless DR leads to atomic fluorine.  We assume a branching ratio of one-half for the formation of atomic fluorine.  

\subsubsection{Photodissociation of HF}

The photoabsorption spectrum of HF (Hitchcock et al.\ 1984; Nee, Suto \& Lee 1985), together with theoretical potential energy curves (e.g. Brown \& Balint-Kurti 2000; see Figure 2) imply that photodissociation is dominated by absorption to the repulsive A$^1\Pi$ state, which has a vertical excitation energy $\sim 84000\,\rm cm^{-1}$.  Theoretical calculations (Brown \& Balint-Kurti 2000) indicate a maximum photodissociation cross-section of $\rm 3.25 \times 10^{-18} \, cm^2$ for HF at a wavelength of 119~nm, values in good agreement with the experimental results of Nee, Suto \& Lee (1985).  The maximum cross-section must still be regarded as somewhat uncertain, however, because experimental results of Hitchcock et al.\ (1984) suggest a value that is larger by a factor 2.

Given the photodissociation cross-section given by Brown \& Balint-Kurti, we obtained a HF photodissociation rate of $1.17 \times 10^{-10}\,\rm s^{-1}\, \chi_{UV}$, where $\chi_{UV}$ is the mean intensity of the radiation field normalized with respect to the 
standard interstellar ultraviolet (ISUV) radiation field of Draine (1978).  For the case of a semi-infinite slab that is illuminated isotropically, we determined how the HF photodissociation rate varies with position using photodissociation region model of Le Bourlot et al.\ (1993), which provides as output the UV radiation field as a function of depth into an interstellar cloud.  We found that the photodissociation rate diminishes as E$_2(2.21 A_V)$, where $A_V$ is the visual extinction in magnitudes behind the slab surface and E$_2$ is the exponential integral of order 2.  The UV dust albedo assumed here was $\omega = 0.32$ and the forward scattering function was 0.73 (Li \& Draine 2001).  

In the deep interiors of clouds with substantial $A_V$, the UV radiation field is dominated by Lyman and Werner band emissions from H$_2$ molecules that have been excited by the secondary electrons ejected by cosmic rays (e.g.\ Prasad \& Tarafdar 1983).  The resulting photodissociation rates have been presented for several molecules of astrophysical interest by Sternberg, Dalgarno \& Lepp (1987; hereafter SDL87) and by Gredel et al.\ (1989).
We have computed the rate of cosmic-ray induced photodissociation for HF, adopting similar methods to those described by Gredel et al.\ (1989).  Here we used the 
H$_2$ UV spectrum plotted in SDL87 (Figure 1) with the more recent excitation yields given by Dalgarno, Yan \& Liu (1999; Table 2).
The result is 87~$\zeta_{\rm cr}/(1-\omega)$, where $\zeta_{\rm cr}$ is the total cosmic ray ionization rate per H$_2$ molecule. 

\subsection{Other reactions}

Although we do not expect them to be significant, for completeness we include the following additional reactions in the chemical network leading to HF:
$$\rm F + CH \rightarrow HF + C \eqno({\rm R13}) $$
$$\rm F + OH \rightarrow HF + O \eqno({\rm R14}) $$
$$\rm F + H_3^+ \rightarrow H_2F^+ + H \eqno({\rm R15}) $$
For reactions (R13) and (R14), we adopt the same rate coefficient assumed for reaction (R2).  As in the case of nitrogen, proton transfer from $\rm H_3^+$ to F is endothermic, but reaction (R15) is exothermic.   Here we adopt the rate coefficient measured for the analogous reaction with N (Scott et al.\ 1997).
  

The rate coefficients for reactions for the entire fluorine reaction network (R1) -- (R15) are summarized in Table 4, together with the rates of photodissociation by the interstellar and cosmic-ray-induced UV radiation fields.

\section{Interstellar cloud model}

We have used a modified version of the photodissociation region 
(PDR) models of Kaufman et al.\ (1999; hereafter K99), Wolfire et al.\ (1990), 
and Tielens \& Hollenbach (1985)
to examine the chemistry of fluorine molecules in diffuse and
dense molecular clouds. In \S 3.1 we discuss results
for a cloud model illuminated from one side and in \S 3.3 
we discuss results for a  cloud model illuminated from two
sides.
The PDR models solve
simultaneously for the thermal balance and chemical equilibrium
abundances of a layer of gas illuminated by a far-ultraviolet 
(6 eV $\le h\nu \le 13.6$ eV; FUV)
radiation field.\footnote{In the K99 models the radiation 
intensity incident on the cloud was normalized to units of the
Habing (1968) field. Here, we adopt a Draine (1978) 
field which has an integrated 6 eV to 13.6 eV intensity that is
1.7 times higher than Habing and denote this field 
strength as $\chi_{UV}=1$. Although Weingartner \& Draine (2001) have recently
advocated using the field derived by Mezger et al.\ (1982) and
Mathis et al.\ (1983) (1.13 times Habing), Parravano et al.\ (2003) found a median 
field closer to that of Draine (1978). Furthermore, 
Wolfire et al.\ (2003) found that a field 1.7 times Habing provides
a good match to the temperature and pressure
of the cold and warm phases of the local ISM.} 

The PDR model for 2-sided illumination is based on the one-sided 
illumination models but modified so as to allow for illumination from both
sides. In the one-sided case, calculation of the chemistry, cooling,
and line-transfer proceeds from the surface to the cloud center in
a single pass since these parameters depend only on the cloud properties
closer to the surface. In the two-sided case, since shielding and
cooling may by important toward both sides of the cloud, an iterative
procedure is required. We first calculate the structure of a cloud
illuminated from one side to a depth of one-half the visual extinction of the
desired two sided model. This provides us with initial estimates of line
optical depths as well as shielding columns of ${\rm H_2}$  and CO. 
In the two-sided
model, a particular point in the cloud sees FUV radiation from both sides and 
cooling radiation
may escape toward either surface; thus we iterate the chemical profiles,
optical depths, and shielding factors.

Modifications to the physics and chemistry assumed in the K99 model have 
been discussed in 
Wolfire et al.\ (2003), who assumed a
higher abundance of PAHs and a  
slower rate of interaction of ions and PAHs.
These changes largely offset each
other with a resulting minor effect on the   
grain heating rates and ion chemistry. 
In addition, we have adopted the results of P\'equignot (1990) 
for the rate coefficient for collisional excitation of O~I by H, and those of McCall et al.\ (2003) for the ${\rm H_3^+}$ dissociative recombination
rate coefficient.

For the case of the one-sided cloud models, we have implemented
an additional modification by using a subroutine
based upon the publically-available\footnote{http://aristote.obspm.fr/MIS/} code from the Meudon group
(Le Bourlot et al.\ 1993; Le Petit et al.\ 2002) 
to treat ${\rm H_2}$ dissociation,
${\rm H_2}$ heating and ${\rm H_2}$ cooling, although we 
retain the ${\rm H_2}$ formation rate of 
$3\times 10^{-17}n_{\rm H}\, n({\rm H})$ cm$^{-3}$ s$^{-1}$ adopted in
K99.  Additional details of our method, test cases, and results,
are discussed in (Kaufman et al.\ 2005).

\subsection{One-sided illumination}

We first consider the case of a constant-density slab that is illuminated from one side by ultraviolet radiation.  As our standard model for a diffuse cloud, we adopt a density of H nuclei $n_H = 10^2$, and a radiation field equal to the average interstellar value given by Draine (1978), $\chi_{UV}=1$.  The illumination is assumed to be isotropic within the hemisphere incident upon the slab surface.  

In Figure 2, we show the temperature profile and molecular abundances plotted as a function of the depth into the cloud (measured in units of visual extinction, $A_V$).  Here, we assume a constant gas-phase fluorine abundance of $1.8 \times 10^{-8}$ relative to H nuclei.  The adopted value is chosen
to fit the abundances $\rm F/O = 
(3.8 \pm 1.7) \times 10^{-5}$ and $\rm O/H = (4.2 \pm 0.2) \times 10^{-4}$ derived by Federman et al.\ (2005) from FUSE observations of atomic fluorine\footnote{Due to the formation of HF, the assumed gas-phase fluorine abundance in our model is $\sim 15\%$ higher than the predicted FI abundance for a cloud with properties
appropriate to the HD 208440 sightline.} along the sightline to HD 208440. 
Similar F/O abundances were inferred from FUSE observations of FI in 
HD 209339A and from the original discovery of FI toward $\delta$ Sco (by {\it Copernicus}; Snow \& York 1981).  The assumed gas-phase fluorine abundance corresponds to roughly 0.6 $\times$ the solar abundance of fluorine. 
The possible effects of fluorine depletion in dense regions are considered in $\S3.4$ below.

The strong tendency towards HF formation is immediately apparent from Figure 2: HF accounts for at least 50\% of the gas-phase fluorine nuclei at all depths $A_V \simgt 0.1$.  Figure 3 compares the rates of various HF formation and destruction processes as a function of depth into the cloud.  At all $A_V$, HF formation is dominated by reaction of F with H$_2$ (reaction R1).   At small $A_V$ ($A_V \simlt 2$ for this particular case), HF destruction is dominated by reaction with C$^+$ (R3) and by photodissociation by ISUV radiation.   These processes become less rapid with increasing $A_V$ as the C$^+$ abundance drops in favor of C and CO and the ISUV field is attenuated by dust absorption.  For intermediate $A_V$ (in the range $\sim 2 - 4$ for this case), reaction with Si$^+$
(R4) becomes dominant.  Once the Si$^+$ density begins to drop at $A_V \simgt 4$, HF destruction is dominated by reaction with H$_3^+$ and He$^+$.  In this regime, the abundances of H$_3^+$ and He$^+$ are controlled by the cosmic ray ionization rate and are essentially independent of $A_V$; this effect provides a ``floor" below which the total HF destruction rate never drops.

\subsection{Parameter study}

We also obtained results for incident radiation with intensities, $\chi_{UV}$, of $10^{-1}$, 1, 10, 100, 10$^3$, and 10$^4$ times the average radiation field given by Draine (1978); and for H nucleus densities, n$_H$, of 10$^{1.5}$, 10$^2$, 10$^{2.5}$, 10$^3$, 10$^{3.5}$, and 10$^4 \rm \, cm^{-3}$. In Figure 4, we present HF abundance profiles for selected cases covering a wide range of assumed density and ISUV field.  Here the abundance plotted is normalized with respect the total gas-phase abundance of fluorine nuclei.  The tendency to HF formation is strikingly robust. In every case we considered, substantial HF abundances are achieved at depths $A_V > 2$ below the cloud surface.  Indeed, the abundance of HF closely mirrors that of H$_2$ (as expected given the fact that HF is formed by direct reaction of H$_2$ with F).  Figure 5 shows the same data presented in Figure 4, now expressed in terms of HF/H$_2$ abundance ratio; its value always lies within an order of magnitude of $3.6 \times 10^{-8}$, the value achieved in the limit of large $A_V$ when H$_2$ and HF are the dominant reservoirs of gas-phase H and F nuclei. 

\subsection{Two-sided illumination}

We have extended the results presented in Figures 4 and 5 to the case where radiation is incident upon both sides of slab of finite thickness.  Here calculations made use of the modifications to the PDR code described at the beginning of \S3.  In Figure 6, solid lines show the total H$_2$ and HF column densities, $N({\rm H_2})$ and $N({\rm HF})$, in a slab with $\chi_{UV} = 1$, $n_{\rm H} = 10^2\rm cm^{-3}$; the results are shown as a function of the total visual extinction through the cloud, $A_V({\rm tot})$.  As expected from Figure 5, the H$_2$ and HF column densities track each other closely.  In Figure 7, we plot $N({\rm HF})$ as a function of $N({\rm H_2})$ for several values of $\chi_{UV}$ and of $n_{\rm H}$.  The dashed line corresponds to $N({\rm HF})/N({\rm H}_2) = 3.6 \times 10^{-8}$, the value obtained in the limit where H$_2$ and HF are the sole gas-phase reservoirs of H and F nuclei.  In Figure 8,  $N({\rm HF})$ is shown as a function of $A_V({\rm tot})$. 

We note that the column densities and visual extinctions plotted in Figures 6 -- 8 all refer to a single slab viewed face-on.
Geometric effects can lead to significant multiplicative factors that affect the measured column densities and extinctions. 
For example, a plane-parallel slab viewed at an inclination angle $\theta$ will show line-of-sight column densities and extinctions that exceed the plotted quantities by a factor sec$\,\theta$.  The presence of multiple clouds on a given sight-line can lead to similar effects.

\subsection{Effects of variable fluorine depletion}

The theoretical results presented in \S3 all assume that the depletion of fluorine nuclei is independent of $A_V$.  The observational data, however, suggest that the fluorine depletion is far greater in dense molecular clouds than in diffuse clouds.  In particular, while a gas-phase fluorine abundance $\sim 0.6 \,\times$ solar has been determined from observations of atomic fluorine in diffuse clouds (see \S 3.1 above), a gas-phase fluorine abundance of only $\sim 0.02 \, \times$ solar has been inferred (Neufeld et al.\ 1998) from far-infrared observations of HF absorption in the dense gas associated with Sgr B2.  Similar behavior has been inferred for the element chlorine (Zmuidzinas et al.\ 1995; Schilke et al.\ 1995) from submillimeter observations of Sgr B2 and and OMC-1. 

We have considered the possible effects of fluorine depletion using the simple model described below.  A similar model has been recently proposed (Hollenbach et al.\ 2005) to explain the freeze-out of interstellar oxygen nuclei to form icy grain mantles. 

We assume that F atoms and HF molecules in the gas-phase accrete onto dust grains at a rate proportional to the density of dust particles (which, for a fixed gas-to-dust ratio, is proportional to the density of H nuclei, $n_{\rm H}$.)   In the case where F atoms are the accreting species, we assume that reaction with an adsorbed H atom leads rapidly to the formation 
that remains bound to the grain mantle\footnote{The probability of retention is unknown, but is probably large (as it is for the grain surface reaction of O + H, for which a large retention probability $\simgt 0.5$ must be invoked to explain the observed abundances of interstellar water ice, e.g.\ Jones \& Williams 1984)}.  Except for small grains of size $\simlt 20\AA$ that are periodically``spike-heated" (Hollenbach et al.\ 2005) to high temperatures ($\sim 100$~K), the HF is assumed to remain frozen onto the grain surface until photodesorbed by ultraviolet radiation.
Thus we assume HF molecules to be removed by photodesorption via the ``direct mechanism" described by Draine and Salpeter (1979), in which excitation of electronic states is followed by ejection from the grain surface with probability, $\epsilon$.   Because the ultraviolet photoabsorption cross-section is dominated by photodissociation, the removal rate for F nuclei in the grain mantle is simply given by $\epsilon\,\zeta_{\rm pd}({\rm HF})$, where $\zeta_{\rm pd}({\rm HF})$ is HF photodissociation rate given in Table 4\footnote{Presumably, the ejection probability $\epsilon$ drops rapidly with depth below the surface of the grain mantle.  Thus, when averaged over all molecules in the mantle, the value of $\epsilon$ becomes a decreasing 
function of mantle thickness once that thickness exceeds a single monolayer.}.  As the HF photodissociation rate drops with increasing extinction, the fraction of fluorine nuclei in the gas-phase diminishes and the abundance of HF ice increases.  Our model is motivated by absorption-line observations of water ice (e.g. Whittet et al.\ 2001), which suggest that significant abundances of water ice are present only in sight-lines along which the extinction exceeds a certain threshold.  

In equilibrium, this simple model implies that the ratio of fluorine nuclei in the gas-phase to HF molecules in the solid phase is given by
$$ R = {n_{\rm F}\,{\rm (gas)} \over n({\rm solid \,\, HF})} = 10^{14}{\rm \,cm^{-3} \, s\,}{\epsilon_{-3} \over k_{-17}}\, {\zeta_{\rm pd}({\rm HF}) \over n_H}, \eqno(5)$$
where $10^{-3} \epsilon_{-3}$ is the probability that the electronic excitation of a solid HF molecule will be followed by ejection from the grain mantle, and $10^{-17}\, k_{-17}\, n_H \rm \, cm^3 \, s^{-1}$ is the accretion rate per F atom or HF molecule in the gas-phase.  Equilibrium is achieved on the accretion timescale, $(3~{\rm Myr} / k_{-17}) ({\rm 10^3} / n_H)$.  

In estimating the ratio $\epsilon_{-3} / k_{-17}$, we turn to the observations of water ice, assuming that the water ice abundance is determined by an analogous balance between photodesorption and accretion. 
From observations of the 3$\,\mu$m water ice absorption feature in the spectra of a sample of stars in the Taurus region, Whittet et al.\ (2001) inferred that a threshold extinction $A_V \sim 3.2$~mag was required for a sight-line to yield a significant water ice abundance.  Requiring that the quantity $10^{14}\,(\epsilon_{-3} / k_{-17})\, \, \zeta_{\rm pd}({\rm H_2O}) / n_H$ should equal unity in a cloud of total extinction $\sim 3.2$~mag (along a diameter), adopting the ${\rm H_2O}$ photodissociation rate $\zeta_{\rm pd}({\rm H_2O})$ derived by Roberge et al.\ (1991), and assuming a typical density $n_{\rm H} \sim 10^3 \rm \,cm^{-3}$ for the clouds probed by the Whittet et al.\ (2001) study, we obtain an estimate of $0.45$ for the ratio $(\epsilon_{-3} / k_{-17})$. Here we are implicitly assuming that the physics governing the freeze-out of HF is similar to that of H$_2$O.  This assumption seems plausible given the similar dipole moments (both $\sim 1.8$~D) 
and melting points (190~K for HF versus 273~K for H$_2$O) of HF and H$_2$, and the fact that the ultraviolet photoabsorption cross-section is dominated by photodissociation in both cases.  We are making an additional approximation by selecting the value of $\epsilon$ for the ``threshold" extinction and neglecting its variation with grain mantle thickness (see footnote 8).

Given the simple model described above, we may estimate the gas-phase abundance of fluorine nuclei as
$$ {n_{\rm F}\,{\rm (gas)} \over n_H} = 1.8 \times 10^{-8} { R \over 1 + R} \eqno(6)$$
where $1.8 \times 10^{-8}$ is the ``diffuse cloud abundance" ($\sim 0.6 \, \times$ solar) obtained in the limit of low density and low extinction. Since the timescale for freeze-out is long compared to the timescale for the gas-phase reactions that set the HF/F ratio in the gas, the effect of freeze-out is simply to lower the overall abundances of gas-phase F and HF but to leave their abundance {\it ratio} unchanged.  

We have recomputed the predicted dependence of $N({\rm HF})$ upon $N({\rm H_2})$, now taking account of variable fluorine depletion with the aid of the equations (5) and (6).  The results for a two-sided PDR with $n_{\rm H} = 10^2$ and $\chi_{UV} = 1$ are represented by the dashed line in Figure 6.  As expected, the predicted column densities of gaseous HF are unchanged for small values of $A_V({\rm tot})$, but approach a maximum value at $A_V({\rm tot}) \sim 3$.  Results for the parameter study are shown in Figures 9 and 10, which are presented in a manner entirely analogous to Figures 7 and 8. 

\section{Discussion}

The results shown in Figures 2 -- 10 show that there is a dramatic tendency towards molecule formation in the chemistry of interstellar fluorine.  In diffuse clouds of small extinction, the predicted HF abundance can even exceed that of CO, even though the gas-phase fluorine abundance is four orders of magnitude smaller than that of carbon.  In the transition from atomic clouds of small $A_V$ to diffuse molecular clouds, hydrogen and fluorine are the first elements to become molecular.  HF is predicted to be the most abundant fluorine-bearing molecule by far; the next most abundant molecule containing fluorine, CF$^+$, is expected to show an abundance
roughly two orders of magnitude smaller than that of HF.  The CF$^+$/HF abundance ratio is given by $k_3 n({\rm C}^+)/k_9 n_e$, where $k_3$ and $k_9$ are the rate coefficients for reactions R3 (reaction of C$^+$ with H$_2$) and R9 (dissociative recombination of CF$^+$). 
In the surface layers of the cloud, where the electron
and C$^+$ abundances are nearly equal, the CF$^+$/HF abundance ratio becomes simply $k_3/k_9 = 0.036 (T/300\,\rm K)^{0.35}$.  In clouds of $A_V \simgt 1$~mag, CF$^+$ column densities $\sim 10^{11}$~cm$^{-2}$ are expected.
In dense PDRs, with $n_H \simgt 10^5$~cm$^{-2}$, the CF$^+$ $J=1$ and $J=2$ level populations will approach LTE
and face-on brightness temperatures $\sim$ 5 and 20 mK km s$^{-1}$ are expected\footnote{given a CF$^+$ dipole moment 
of 1.04~Debye, the value obtained in the multireference configuration interaction (MRCI) calculations of Peterson et al.\ (1989)}
respectively
for the 102.6~GHz $J=1-0$ and 205.2~GHz $J=2-1$ transitions.  These lines are potentially detectable from ground-based observatories, particularly in nearly edge-on PDRs where the source geometry enhances the brightness of optically-thin lines.

The large predicted abundances for HF provide a strong motivation for future absorption-line observations  using airborne or satellite observatories.  Such observations are expected to lead to the widespread detection of interstellar HF, even when account is taken of the effects of HF freeze-out (\S4.1 above).  
With its small moment of inertia, HF has a extremely large rotational constant, $B = 20.54$ cm$^{-1}$, which places all of its pure rotational transitions at far-infrared wavelengths that are inaccessible to ground-based observatories. To date, observations of interstellar HF have been limited to a single transition in a single source; the $J=2-1$ transition at $121.7\rm \, \mu m$ was detected by the Infrared Space Observatory (ISO) in absorption towards the source Sgr B2 (NZSP).  Because the HF $J=1-0$ transition possesses an extremely large critical density (see \S 2.3.1 above), significant populations of HF $J=1$ are typically achieved only in interstellar clouds that are strongly irradiated by far-infrared continuum emission; Sgr B2 is the classic example.  The need for strong radiative pumping to populate HF $J=1$ severely limits the utility of HF $J=2-1$ absorption as a probe of interstellar hydrogen fluoride.  

While the HF $J=1-0$ transition at 243.2~$\mu$m lies outside the spectral range probed by the ISO spectrometers, it will become observable with the advent of the Stratospheric Observatory for Infrared Astronomy (SOFIA) and the Herschel Space Observatory (HSO).  In particular, the CASIMIR instrument on SOFIA and the HIFI instrument on HSO will yield high resolution heterodyne spectroscopy in the relevant spectral region.  These instruments will provide an absorption line probe of hydrogen fluoride in its ground rotational state.  

Indeed, the HF $J=1-0$ transition promises to yield an extremely sensitive probe of diffuse molecular gas along the line of sight to far-infrared infrared continuum sources.   For a cloud with a HF column density $N({\rm HF}) = N_{12}\,10^{12}\, \rm cm^{-2}$ and a Doppler parameter $b = 10^5\, b_5$~cm~s$^{-1}$, the line-center optical depth is $\tau_0 = 0.35 N_{12}/b_5$.  Thus typical interstellar clouds with HF column densities as small as $\,10^{12}\, \rm cm^{-2}$ should be routinely observable with CASIMIR and HIFI, and HF column densities as small as $\,10^{11}\, \rm cm^{-2}$ are potentially detectable in spectra of high signal-to-noise ratio.

Referring to Figure 6, we find that HF is potentially detectable in diffuse interstellar gas clouds with 
H$_2$ column densities smaller than $\rm 10^{19} cm^{-2}$.  To date, the only molecules detected in clouds of such small size are H$_2$ and HD, observed in absorption toward hot background stars at UV 
wavelengths\footnote{HD is also detectable at far-IR wavelengths via its 112$\mu$m $J=1-0$ transition, which was detected using ISO (Caux et al.\ 2002) and will be observable using the GREAT instrument on SOFIA.  Unfortunately, because the HD dipole moment is so small, the optical depth is only $3 \times 10^{-20} N({\rm HD}) / b_5$.  Thus, far-IR HD absorption will be detectable only along sight-lines with $A_V$ of at least 10~mag, even in spectra of high signal-to-noise ratio.}.  This raises the interesting and (at least to us) surprising possibility that spectroscopic observations of far-infrared continuum sources with SOFIA and HSO will reveal a component of foreground molecular gas that is 
observed primarily by means of its HF absorption lines.  Such observations will also be sensitive to HF emission that is intrinsic to the warm, dense background sources themselves.  

HF also possesses far-ultraviolet absorption lines that could be searched for in archival stellar spectra obtained with satellite observatories that show diffuse cloud absorption.  The most favorable target line is likely the $\rm C ^1\Pi \gets X ^1\Sigma \,\,\,\ v= 0 - 0 \,\,\,\ R(0)$ transition at $951.266 \AA$ (Tashiro et al.\ 1989) -- which has an estimated oscillator strength $f=0.039$ (Hitchcock et al.\ 1984) -- although very high spectral resolution would be needed to resolve it from the nearby H$_2$ $951.681 \AA$ and NI $951.295 \AA$ features.  

Hydrogen fluoride is also a potential probe of molecular material at high redshifts.  To date, such material has been studied (1) by means of CO rotational emission lines; and (2) by absorption line spectroscopy of background QSOs at UV/optical and at radio/millimeter wavelengths.  Thus far, H$_2$ has been detected directly at UV and optical wavelengths in 8 damped Lyman alpha systems at redshifts ranging from 2.0 to 3.0, by means of the Lyman and Werner absorption lines (Ledoux, Petitjean, \& Srianand 2003), and references therein); the observed molecular fractions, $f = 2N({\rm H}_2)/N_H$, are typically less than 0.03 and the H$_2$ column densities smaller than $2 \times 10^{18} \, \rm cm^{-2}$. At millimeter wavelengths, foreground molecular gas has been detected in absorption towards 4 high-redshift radio sources at redshifts in the range z = 0.25 -- 0.89 (Wiklind \& Combes 1999; Curran et al.\ 2004).  For absorbing material detected at radio and millimeter wavelengths, the visual extinction along the sight-line is typically large and the molecular fraction is $\sim 0.3 - 1.0$.  The list of detected molecules includes CO, HCO$^+$, HCN, H$_2$O, and OH.  

HF observations are a potentially valuable bridge between the UV/optical observations capable of probing low column density systems and the radio/millimeter-wavelength observations that probe clouds of high extinction and large molecular fraction.  At redshifts beyond $z \sim 0.3$, HF is potentially detectable from ground-based observatories in several atmospheric transmission windows.     
Because the HF $J=1-0$ rest frequency is considerably higher than that of the CO, HCO$^+$, HCN, H$_2$O, and OH transitions observed to date, and because the continuum fluxes and receiver sensitivities decline with frequency, the large collecting area to be provided by ALMA will likely be needed to make significant progress in high-z observations of HF. In Figure 11, we show the model atmospheric transmission (Pardo, Cernicharo \& Serabyn 2001) at the redshifted wavelength of HF $J=1-0$,  as a function of the redshift; results apply to the zenith transmission under favorable conditions (0.4~mm precipitable water vapor) at the proposed Chajnantor ALMA site.

\begin{acknowledgments}

We are indebted to C.~Zhu for providing us with unpublished rate coefficients for the reaction $\rm F + H_2 \rightarrow HF + H$ over the critical temperature range 10 -- 50~K.  We are very grateful to the referee for several very helpful comments about the manuscript -- and, in particular, for pointing out that CF$^+$ may be detectable through ground-based observations at millimeter wavelengths. It is a pleasure also to acknowledge valuable discussions with A.~Dalgarno, M.~Kaufman, D.~Hollenbach, P.~Sonnentrucker, P.~Dagdigian, and M.~Alexander. D.A.N.\ gratefully acknowledges the support of a grant from NASA's Long Term Space Astrophysics (LTSA) Research Program.  M.G.W.\ is supported in part by NASA LTSA grant NAG 5-9271.

\end{acknowledgments}

\begin{deluxetable}{lrrlr}
\tablewidth{0pt}
\tablecaption{Enthalpies of formation of hydrides and fluorides\tablenotemark{1}}
\tablehead{
\multicolumn{2}{c}{Hydrides} && \multicolumn{2}{c}{Fluorides} \\
\cline{1-2} \cline{4-5}  
\colhead{\phantom{00000000}} & \colhead{(kJ mol$^{-1}$)}&\colhead{\phantom{00000000}}&\colhead{\phantom{00000000}} & \colhead{(kJ mol$^{-1}$)}}

\startdata

{\bf HF}& {\bf --571} &&   CF$^+$  &  --748\\
H$_2$ &  --436      &&   SiF$^+$ & --608 \\
OH    &  --428      &&   {\bf HF}      & {\bf--571} \\
CH$^+$&  --400      &&   SiF     &  --549\\
SH    &  --356      &&   CF      &  --541\\
SH$^+$&  --350      &&   SF$^+$   & --369 \\
CH    &  --341      &&   SF      &  --344\\
SiH$^+$& --315       &&   NF      & --303 \\
NH    &  --314      &&   OF      &  --220\\ 
SiH   &  --291      &&   F$_2$   &  --159\\

\enddata
\tablenotetext{1} {Enthalpies computed relative to the enthalpies of formation of the constituent atoms, using the values recommended by the NIST Chemistry Web book; the tabulated quantity is therefore $\Delta_f H^0_{\rm gas} ({\rm XY}) - \Delta_f H^0_{\rm gas} ({\rm X}) - \Delta_f H^0_{\rm gas} ({\rm Y})$}.
\end{deluxetable}

\clearpage

\begin{deluxetable}{lccccc}
\tablewidth{450pt}
\tablecaption{Standard heats of reaction
for possible HF-destroying reactions\tablenotemark{1}}
\tablehead{
& \multicolumn{5}{c}{Element, X}\\
\cline{2-6} 
\colhead{Reaction} & \colhead{Carbon}&\colhead{Nitrogen}&\colhead{Oxygen} & \colhead{Silicon} & \colhead{Sulfur}\\}
\startdata
$\rm HF + X \rightarrow XF + H$   & 30\tablenotemark{2}  & 268  & 351 & 21 & 227 \\
$\rm HF + X^+ \rightarrow XF^+ + H$ &{\bf --178}\tablenotemark{3} & 33 &
 269 & {\bf --38} & 202\\
$\rm HF + XH \rightarrow XF + H_2$  & {\bf --66} & 146 & 343 & {\bf --123} & 147\\
$\rm HF + XH^+ \rightarrow XF^+ + H_2$ \phantom{00000}  & {\bf --213} & 30 & 319 & {\bf --159} & 116 \\

\enddata
\tablenotetext{1}{Values computed using the enthalpies of formation recommended by the NIST Chemistry Web book.  For ion-neutral reactions, the ionization potentials recommended by the NIST Chemistry Web book were adopted, and any differences between the specific heat capacity of each ion and its corresponding neutral were neglected.}
\tablenotetext{2}{In units of kJ mol$^{-1}$}
\tablenotetext{3}{Exothermic reactions are flagged by boldface type}
\end{deluxetable}

\clearpage

\begin{deluxetable}{lr}
\tablewidth{0pt}
\tablecaption{Proton affinities}
\tablehead{\colhead{\phantom{00000000}} & \colhead{(kJ mol$^{-1}$)}\\}
\startdata
$\rm F$   & 340\\
$\rm H_2$   & 422\\
$\rm HF$   & 484\\
$\rm N_2$   & 494\\
$\rm CO$   & 594\\
$\rm H_2O$   & 691\\
\enddata
\end{deluxetable}

\clearpage

\begin{deluxetable}{ll}
\tablewidth{0pt}
\tablecaption{Reaction list for F-bearing species}
\tablehead{\colhead{Reaction} & Rate/rate coefficient \\}
\startdata
$\rm F + H_2 \rightarrow HF + H  \phantom{00}$ & $1.0 \times \rm 10^{-10}\,[\,\exp\,(-450\,K/{\it T})+ 0.078\,exp\,(-80\,K/{\it T})$ \\
& $\phantom{1.4 \times \rm 10^{-10}\,} \rm + 0.0155\,exp\,(-10\,K/{\it T})] 
\, cm^3 \,s^{-1}$\\
$\rm F + H_2O \rightarrow HF + OH  $ & $1.6 \times \rm 10^{-10} 
\, cm^3 \,s^{-1}$\\
$\rm C^+ + HF \rightarrow CF^+ + H  $ & $7.2 \times 10^{-9} \, (T/300 \rm \, K)^{-0.15}
\, cm^3 \,s^{-1}$\\
$\rm Si^+ + HF \rightarrow SiF^+ + H $ & $5.7 \times 10^{-9} \, (T/300 \rm \, K)^{-0.15}
\, cm^3 \,s^{-1}$\\
$\rm H_3^+ + HF \rightarrow H_2 + H_2F^+ $ & $1.2 \times 10^{-8} \, (T/300 \rm \, K)^{-0.15}
\, cm^3 \,s^{-1}$\\
$\rm He^+ + HF \rightarrow H + F^+ + He \phantom{000000}$ & $1.1 \times 10^{-8} \, (T/300 \rm \, K)^{-0.15}
\, cm^3 \,s^{-1}$\\
$\rm F^+ + H_2 \rightarrow HF^+ + H $ & $1.0 \times \rm 10^{-9} 
\, cm^3 \,s^{-1}$\\
$\rm HF^+ + H_2 \rightarrow H_2F^+ + H $ & $1.3 \times \rm 10^{-9} 
\, cm^3 \,s^{-1}$\\
$\rm CF^+ + e \rightarrow C + F $ & $2.0 \times 10^{-7} \, (T/300 \rm \, K)^{-0.5}
\, cm^3 \,s^{-1}$\\
$\rm SiF^+ + e \rightarrow Si + F $ & $2.0 \times 10^{-7}\, (T/300 \rm \, K)^{-0.5}
\, cm^3 \,s^{-1}$\\
$\rm HF^+ + e \rightarrow H + F $ & $2.0 \times 10^{-7}\, (T/300 \rm \, K)^{-0.5}
\, cm^3 \,s^{-1}$\\
$\rm H_2F^+ + e \rightarrow HF + H $ & $3.5 \times 10^{-7}\, (T/300 \rm\, K)^{-0.5}
\, cm^3 \,s^{-1}$\\
$\rm H_2F^+ + e \rightarrow F + products$ & $3.5 \times 10^{-7}\, (T/300 \rm\, K)^{-0.5}
\, cm^3 \,s^{-1}$\\
$\rm F + CH \rightarrow HF + C$ & $1.6 \times \rm 10^{-10} 
\, cm^3 \,s^{-1}$\\
$\rm F + OH \rightarrow HF + O$ & $1.6 \times \rm 10^{-10} 
\, cm^3 \,s^{-1}$\\
$\rm F + H_3^+ \rightarrow H_2F^+ + H$ & $4.8 \times \rm 10^{-10} 
\, cm^3 \,s^{-1}$\\
$\rm HF + h\nu \rightarrow H + F $ \phantom{0000000} & $\,1.17 \times 10^{-10}\,\chi_{UV} [\,{1 \over 2}E_2(2.21 A_V) + {1 \over 2}E_2(2.21 [A_{V,{\rm tot}}-A_V])]\,{\rm s}^{-1}\, $ \\
& + $\, 87 \, \zeta_{cr}\, /(1 - \omega) $ \\
\enddata
\end{deluxetable}
\clearpage

\begin{figure}
\plotone{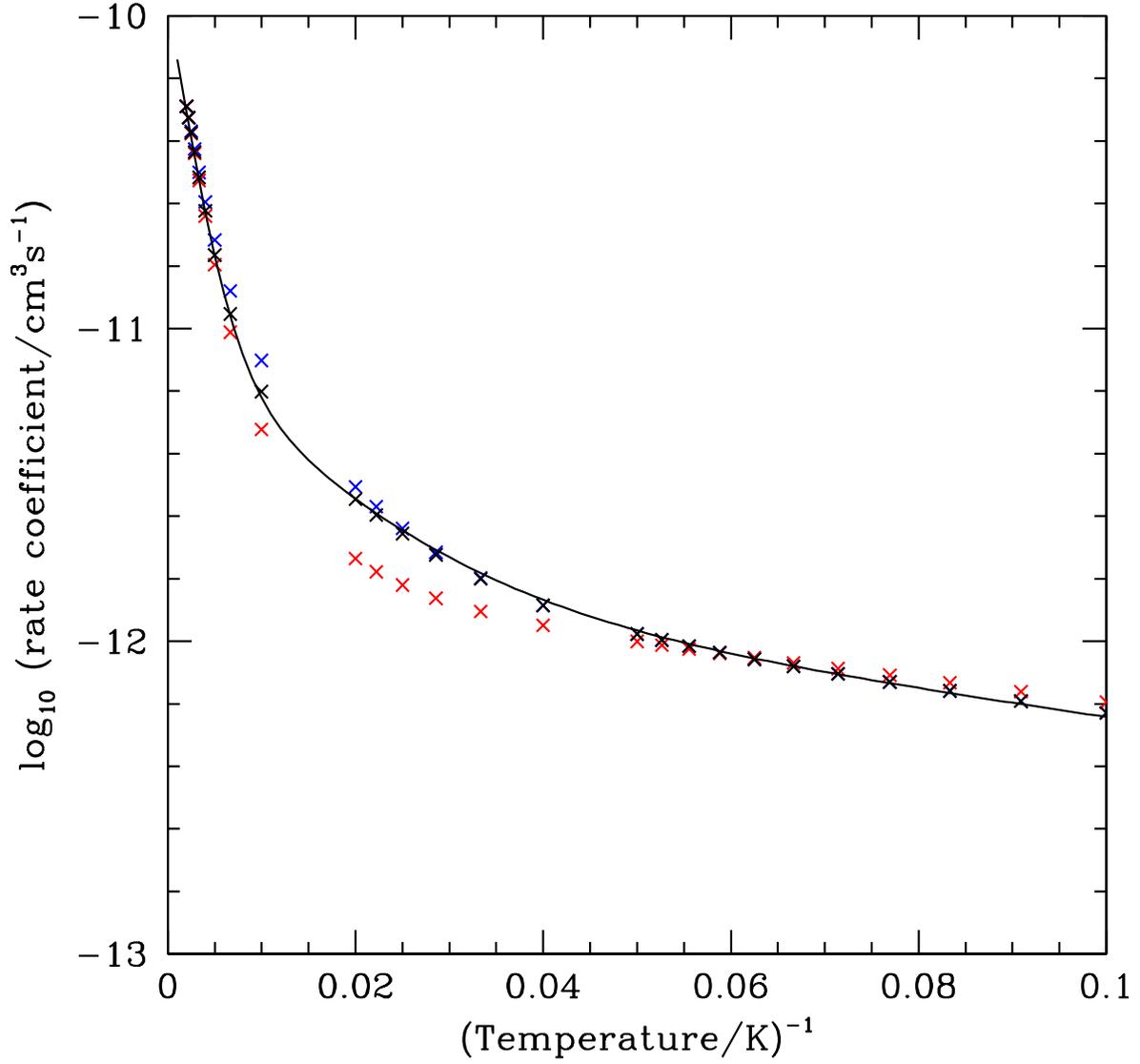}
\figcaption{Rate coefficients computed by Zhu et al.\ for the reaction of F\,($^2P_{3/2}$) with H$_2$ $J=0$ (red crosses) and $J=1$ (blue crosses). 
Black crosses show a weighted average applicable when the H$_2$ populations are in local thermodynamic 
equilibrium, and the solid black curve shows our adopted fit to that average (see Table 4), which is accurate to better than $5\%$ throughout the temperature range plotted.  }
\end{figure}

\begin{figure}
\plotone{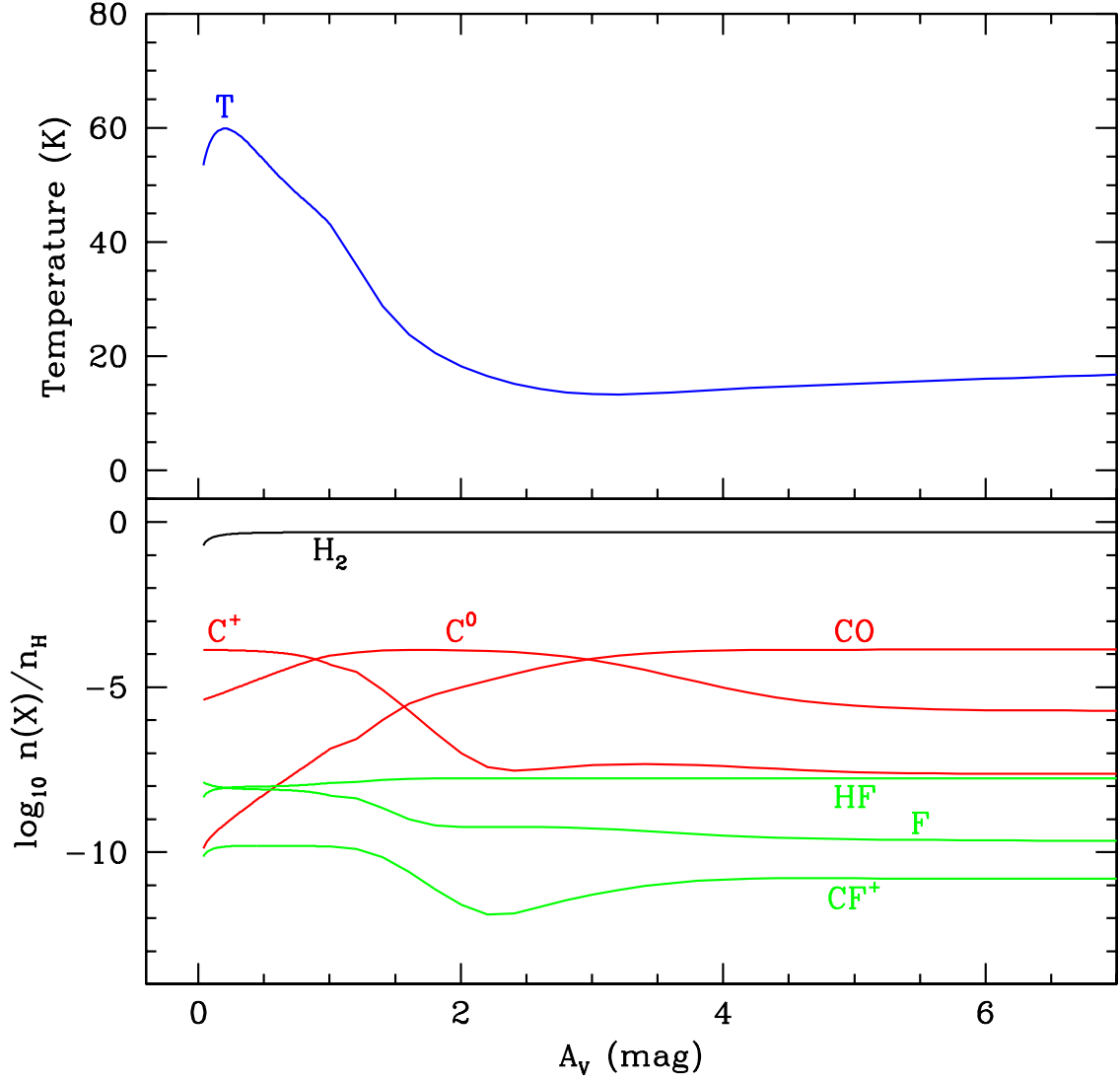}
\figcaption{Gas temperature and abundances in a one-sided PDR with $n_{\rm H} = 10^2 \rm \, cm^{-3}$ and $\chi_{UV} = 1$.}
\end{figure}

\begin{figure}
\plotone{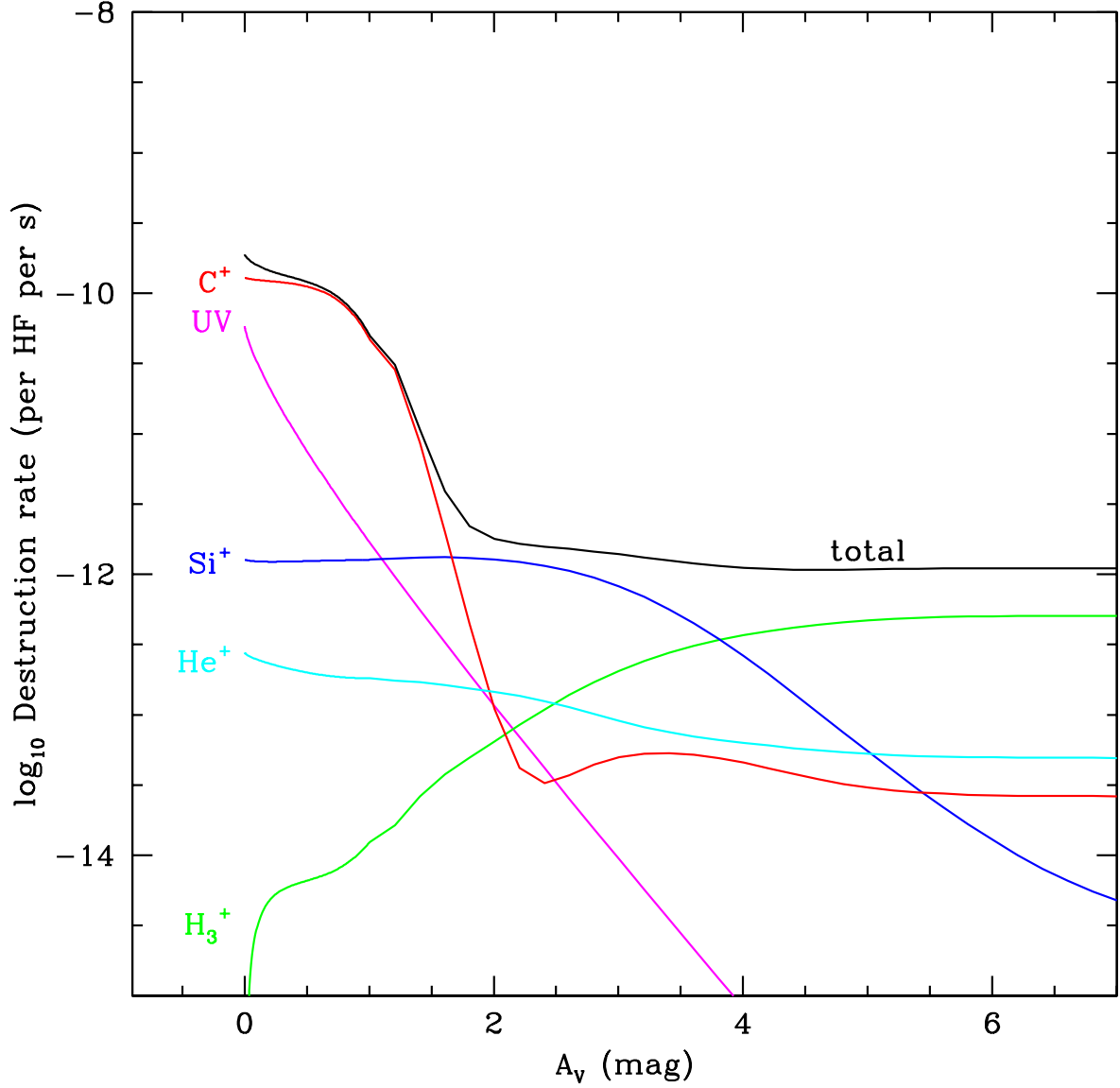}
\figcaption{HF destruction rates in a one-sided PDR with $n_{\rm H} = 10^2 \rm \, cm^{-3}$, $\chi_{UV} = 1$.  Destruction rates are shown for photodissociation by ISUV (magenta curve) and for reactions with C$^+$ (red), Si$^+$ (blue), He$^+$ (cyan) and H$_3^+$ (green).  The black curve shows the total HF destruction rate.}
\end{figure}

\begin{figure}
\plotone{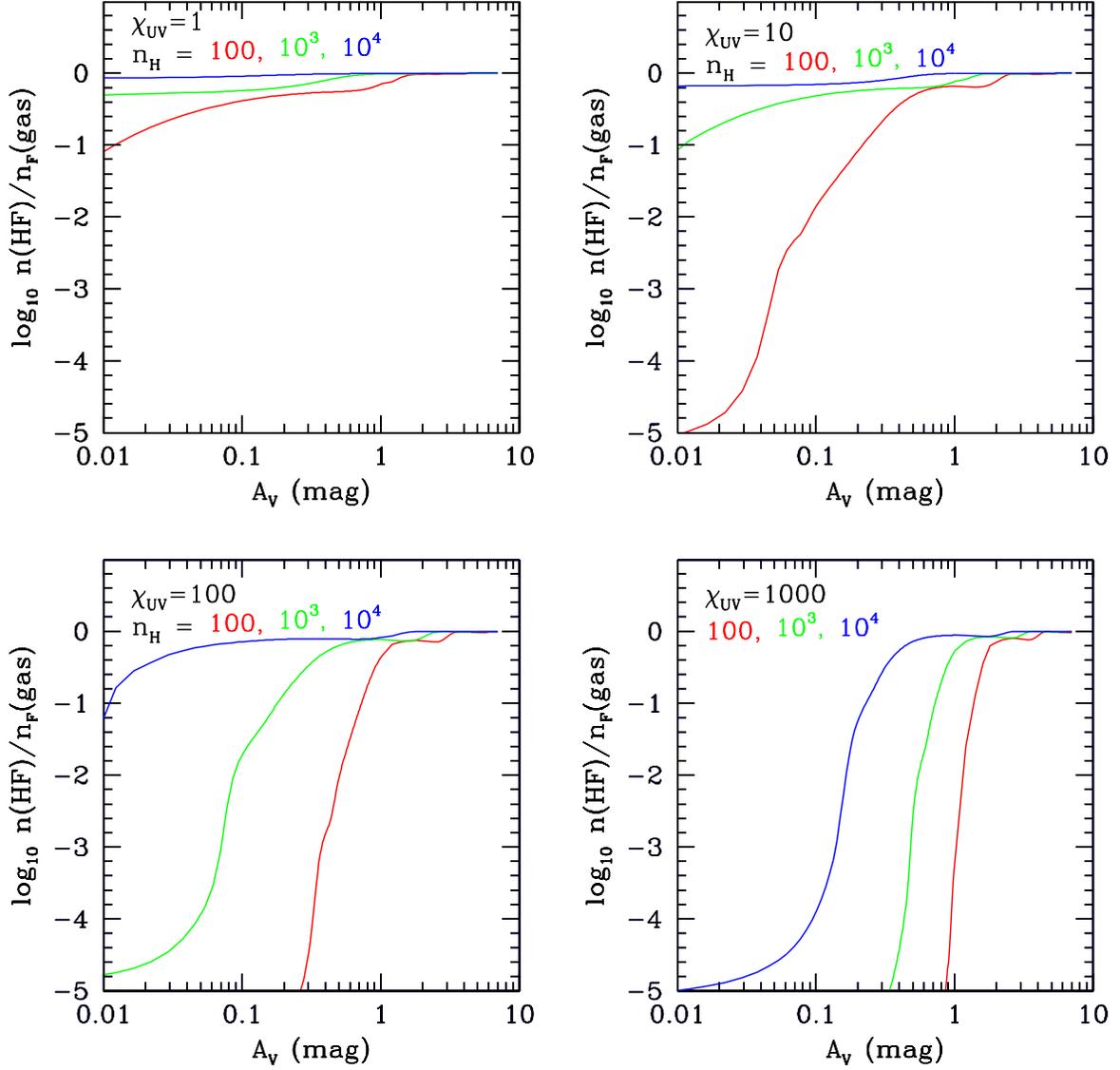}
\figcaption{HF abundance profiles, $n({\rm HF})/n_{\rm F} \rm{(gas)}$, for one-sided PDRs with $n_{\rm H} = 10^2$, $10^3$, and $10^4 \rm \, cm^{-3}$ irradiated by ISUV fields of 
$\chi_{UV} = 1$, 10, 100, and 1000.}
\end{figure}

\begin{figure}
\plotone{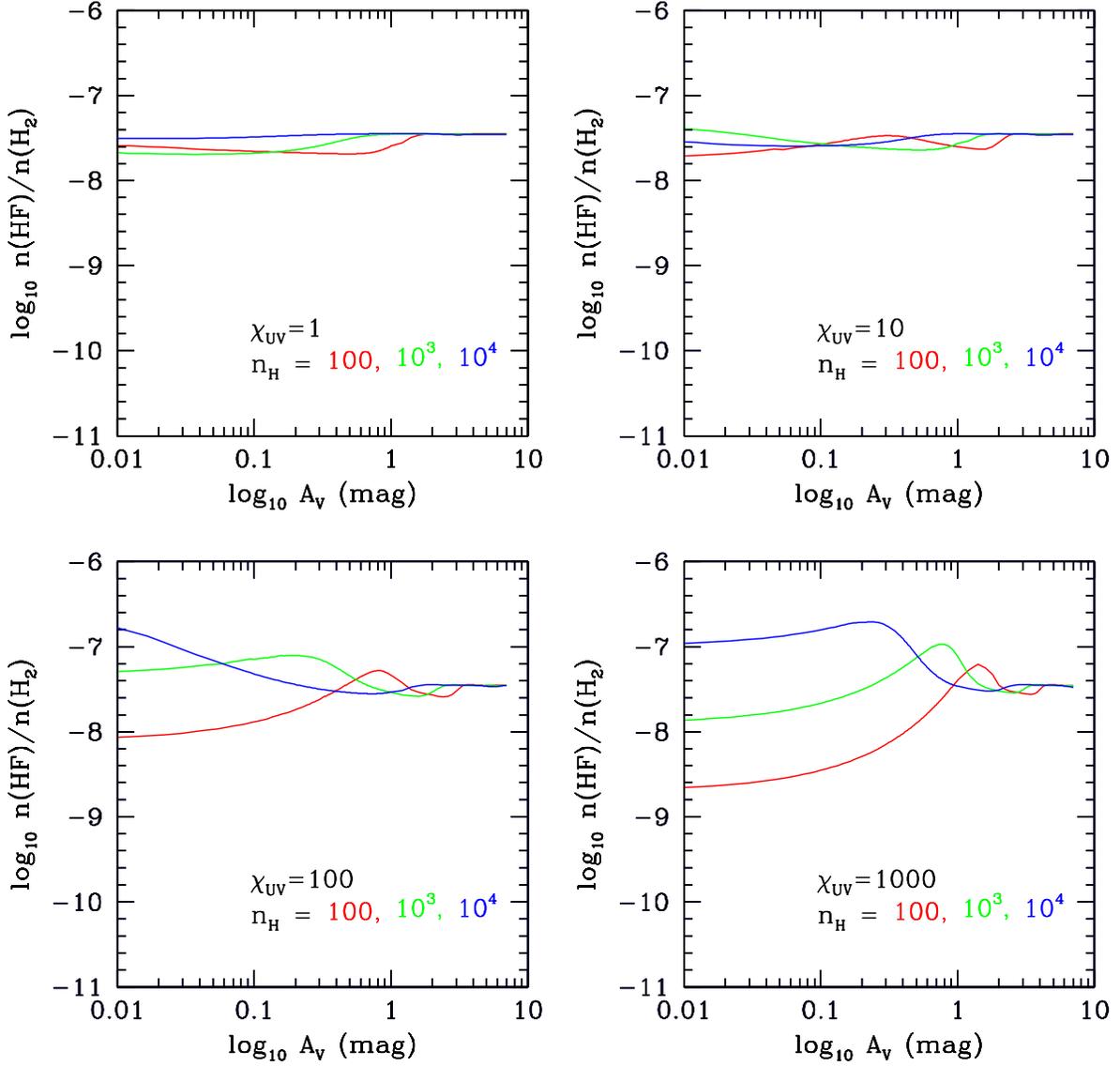}
\figcaption{HF abundance profiles, $n({\rm HF})/n{\rm H}_2$, for one-sided PDRs with $n_{\rm H} = 10^2$, $10^3$, and $10^4 \rm \, cm^{-3}$ irradiated by ISUV fields of $\chi_{UV} = 1$, 10, 100, and 1000.  (Same as Fig.~4, but now with the HF abundance expressed relative to H$_2$ instead of gas-phase fluorine nuclei.)}
\end{figure}

\begin{figure}
\plotone{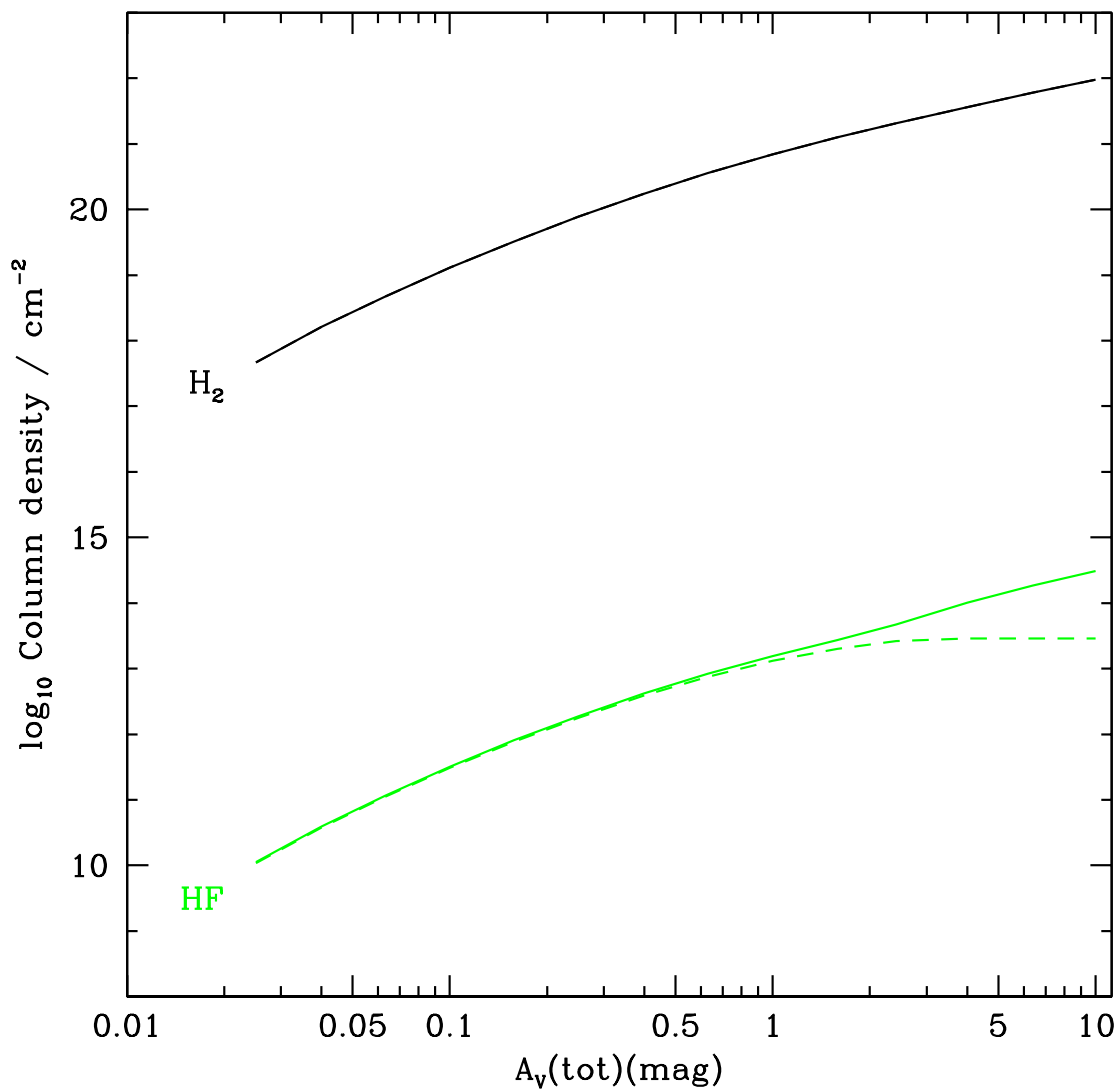}
\figcaption{HF and H$_2$ column densities, for a two-sided PDR with $n_{\rm H} = 10^2$ and $\chi_{UV} = 1$.  Results are plotted as a function of total visual extinction through the cloud, $A_V{\rm (tot)}$.  Solid curves: results for fixed fluorine depletion; dashed curve: results for variable fluorine depletion (i.e.\ with freeze-out included; see \S 3.4)}
\end{figure}

\begin{figure}
\plotone{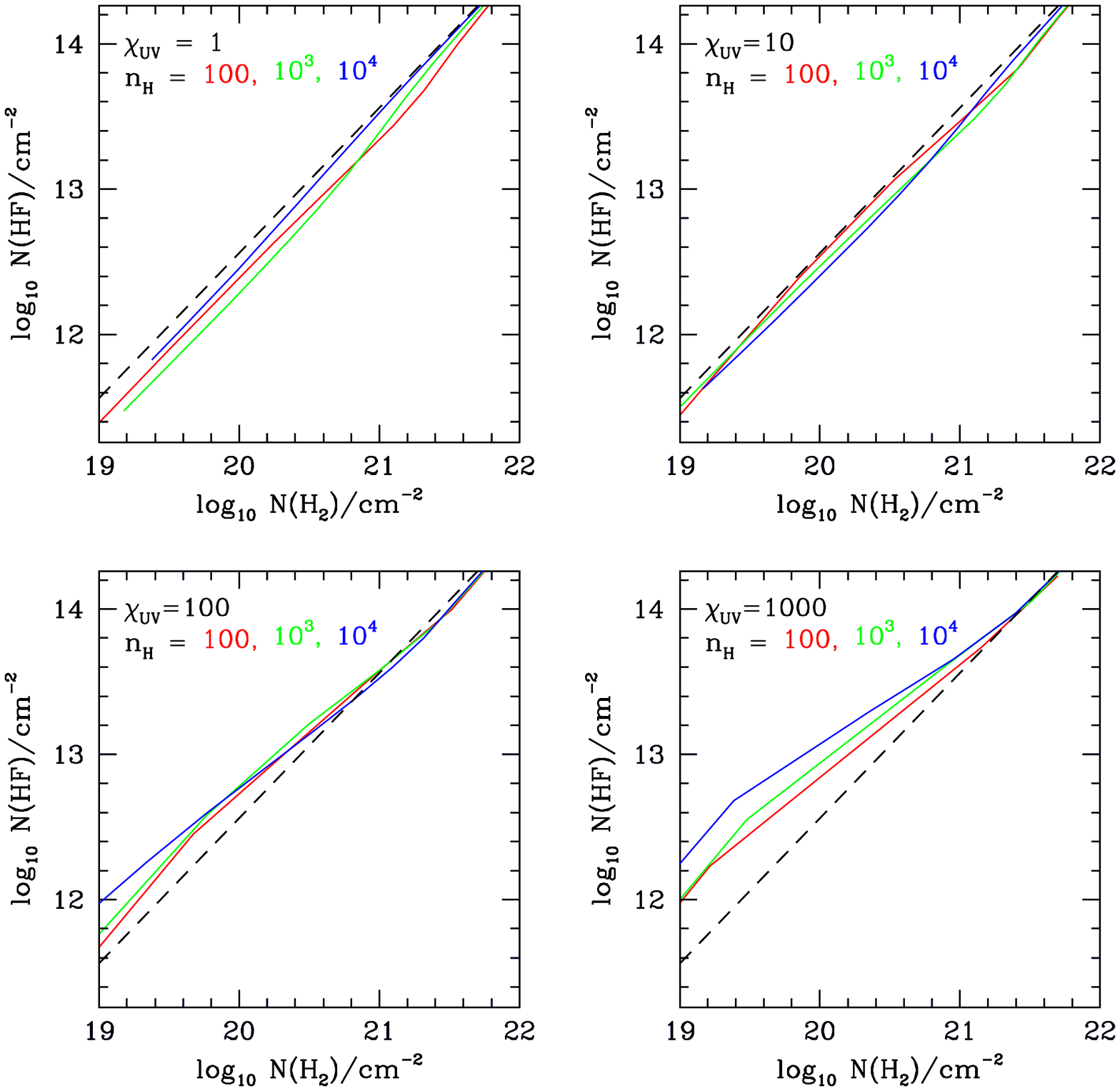}
\figcaption{HF column densities versus H$_2$ column density, for two-sided PDRs with $n_{\rm H} = 10^2$, $10^3$, and $10^4 \rm \, cm^{-3}$ irradiated by ISUV fields of $\chi_{UV} = 1$, 10, 100, and 1000.  The dashed line corresponds to $N({\rm HF})/N({\rm H}_2) = 3.6 \times 10^{-8}$, the value obtained in the limit where H$_2$ and HF are the sole gas-phase reservoirs of H and F nuclei. Results are presented for fixed fluorine depletion.}
\end{figure}

\begin{figure}
\plotone{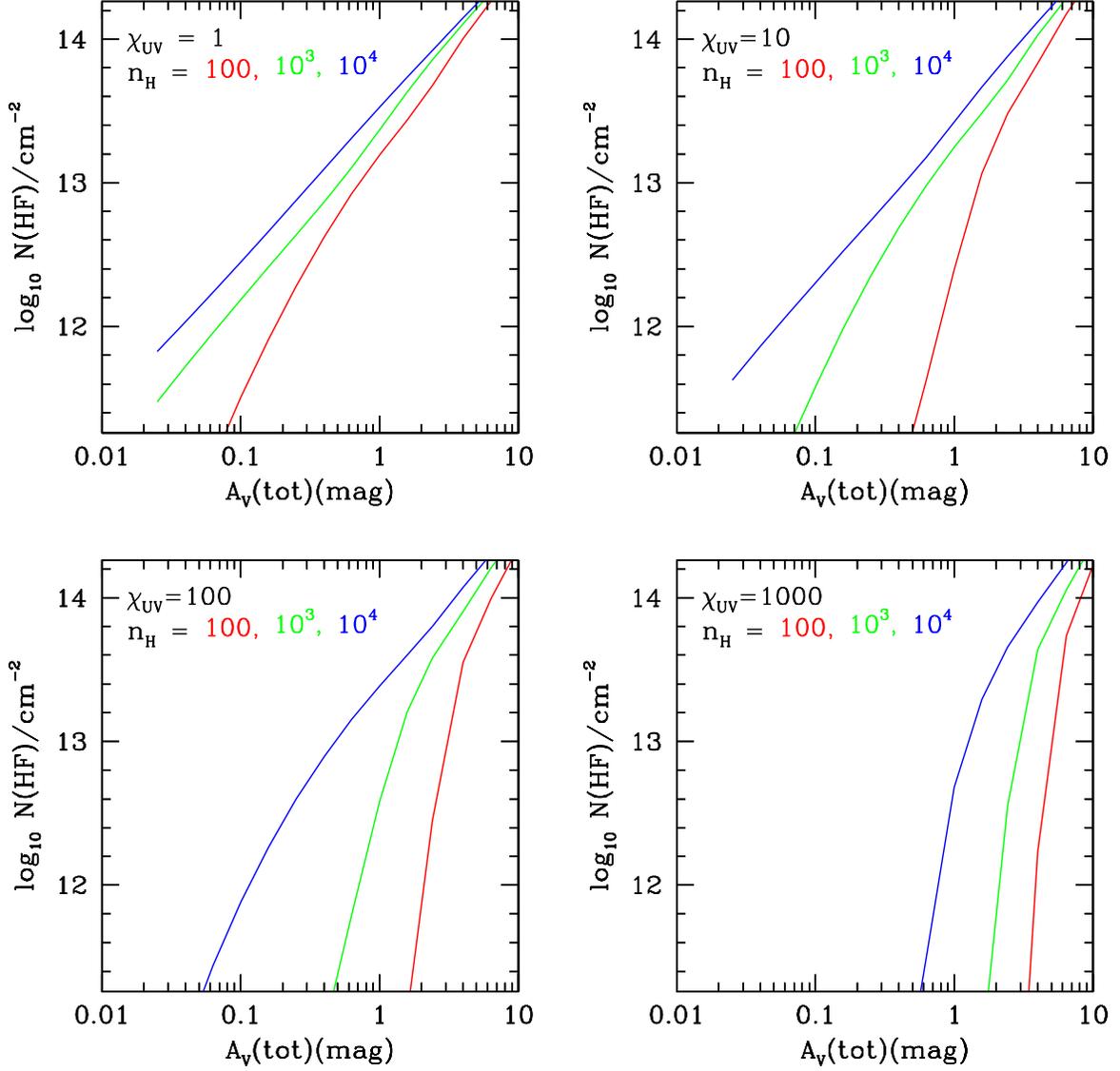}
\figcaption{HF column densities versus total visual extinction, for two-sided PDRs with $n_{\rm H} = 10^2$, $10^3$, and $10^4 \rm \, cm^{-3}$ irradiated by ISUV fields of $\chi_{UV} = 1$, 10, 100, and 1000.  Results are presented for fixed fluorine depletion.}
\end{figure}

\begin{figure}
\plotone{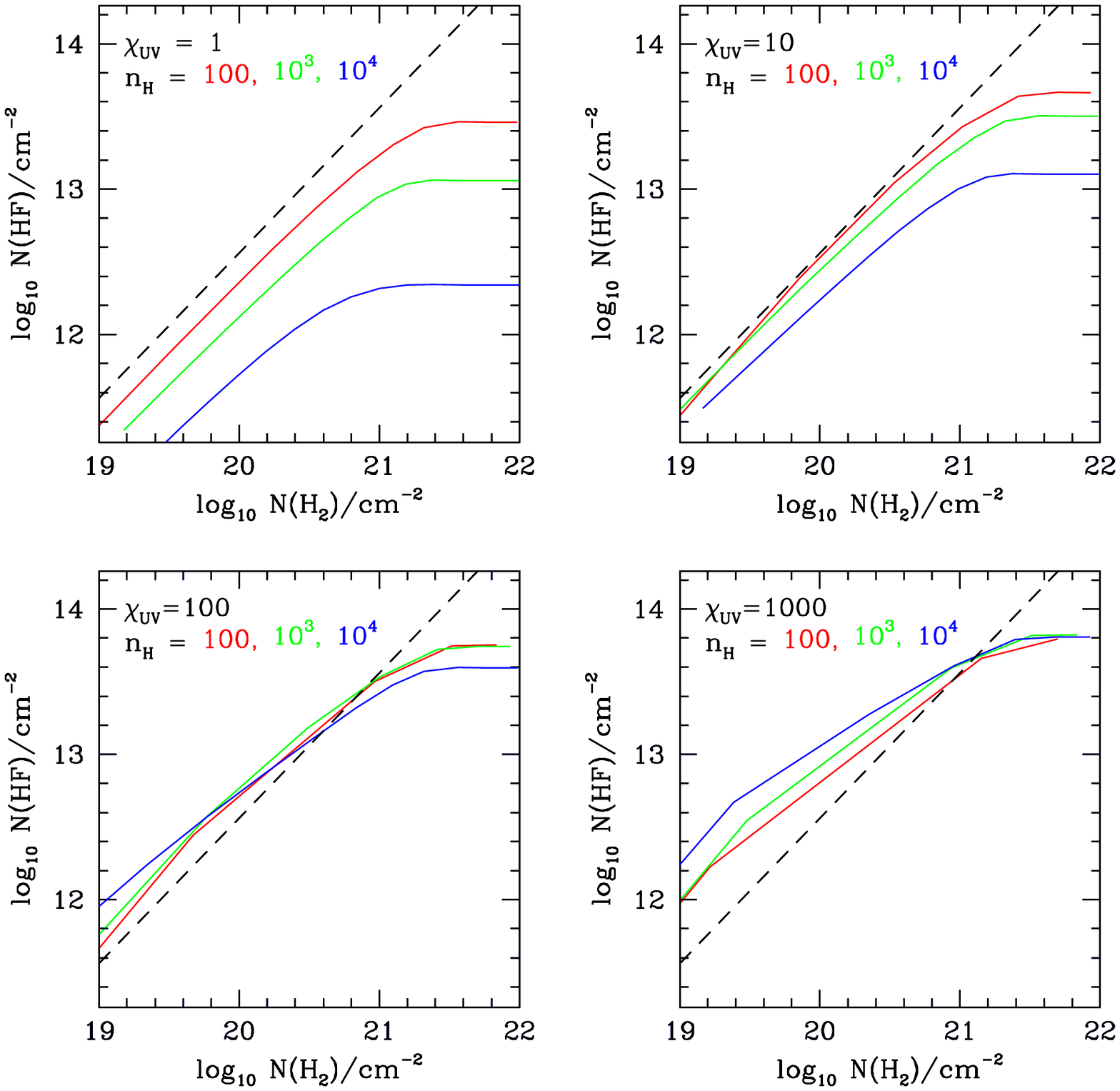}
\figcaption{HF column densities versus H$_2$ column density, for two-sided PDRs with $n_{\rm H} = 10^2$, $10^3$, and $10^4 \rm \, cm^{-3}$ irradiated by ISUV fields of $\chi_{UV} = 1$, 10, 100, and 1000.  The dashed line corresponds to $N({\rm HF})/N({\rm H}_2) = 3.6 \times 10^{-8}$, the value obtained in the limit where H$_2$ and HF are the sole gas-phase reservoirs of H and F nuclei. Results are presented for variable fluorine depletion (i.e.\ with freeze-out included; see \S 3.4)}
\end{figure}

\begin{figure}
\plotone{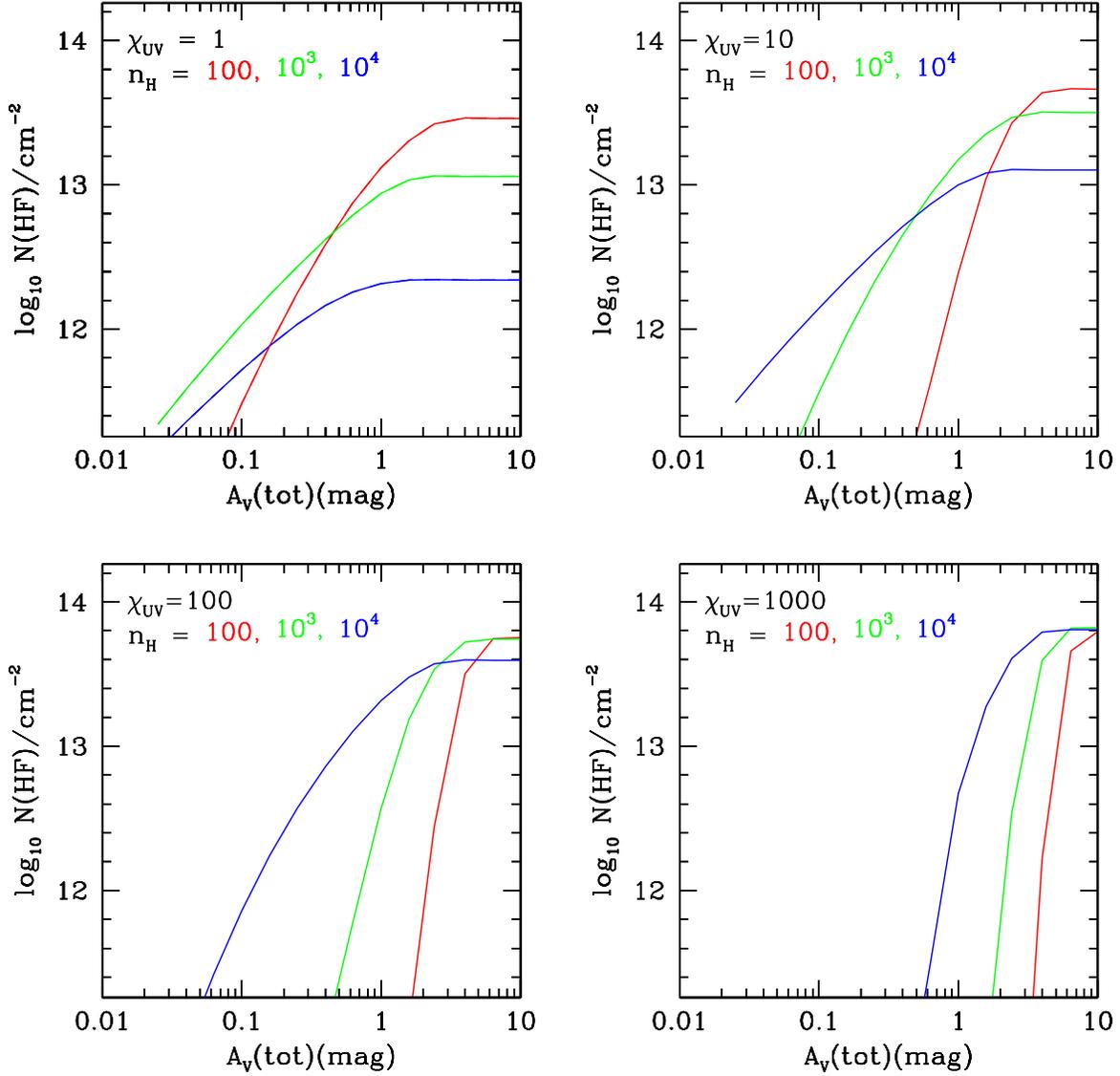}
\figcaption{HF column densities versus total visual extinction, for two-sided PDRs with $n_{\rm H} = 10^2$, $10^3$, and $10^4 \rm \, cm^{-3}$ irradiated by ISUV fields of $\chi_{UV} = 1$, 10, 100, and 1000.  Results are presented for variable fluorine depletion (i.e.\ with freeze-out included; see \S 3.4)}
\end{figure}

\begin{figure}
\plotone{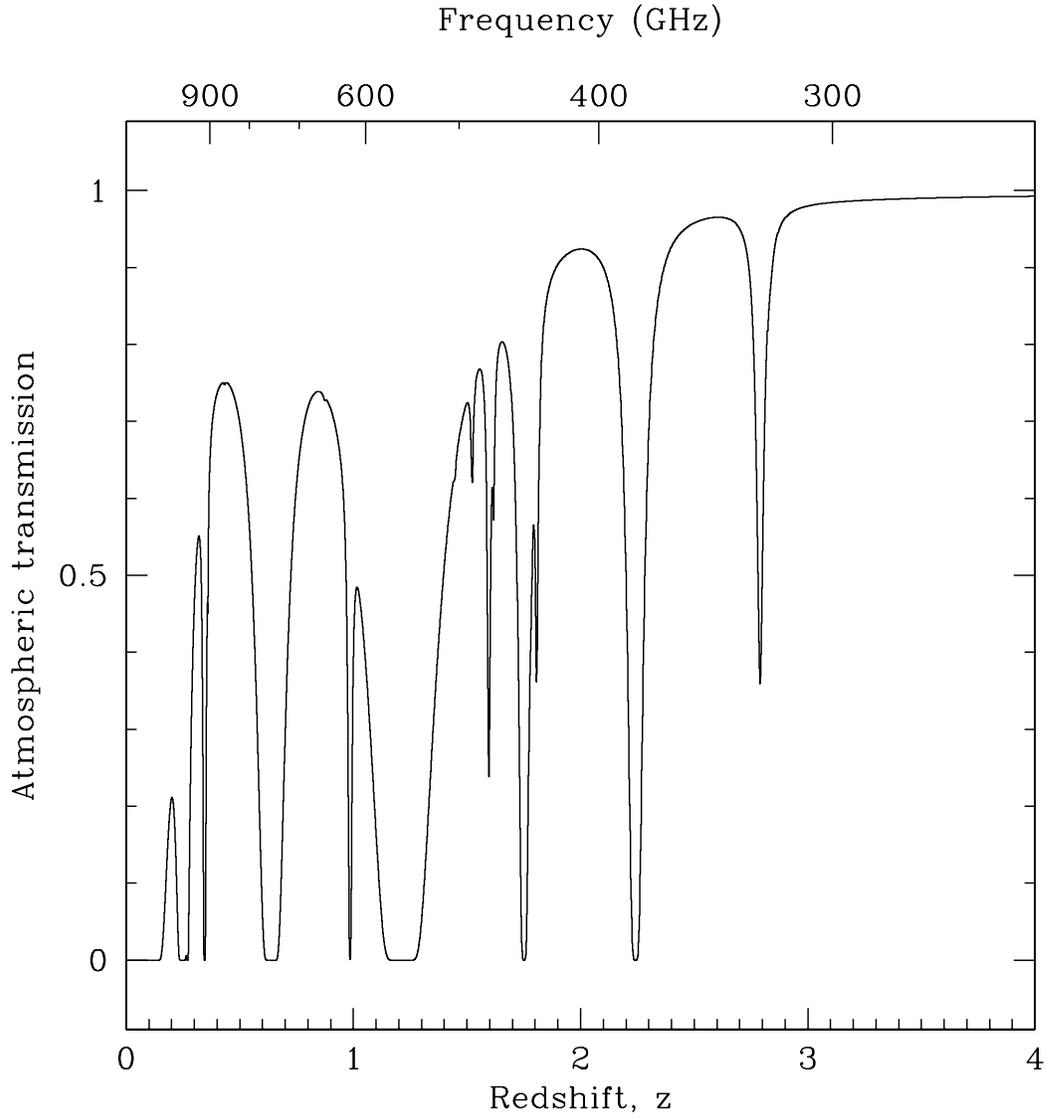}
\figcaption{Model atmospheric transmission (Pardo, Cernicharo \& Serabyn 2001) at the redshifted wavelength of HF $J=1-0$, as a function of the redshift; results apply to the zenith transmission under favorable conditions (0.4~mm precipitable water vapor) at the proposed Chajnantor ALMA site.  The line rest frequency is 1232.476 GHz}
\end{figure}

\begin{thebibliography}{}

\bibitem[Adams et al. (1985)]{ada85} Adams, N.~G., Smith, D., and Clary, D.~C.\ 1985, ApJ, 296, L31

\bibitem[Balakrishnan & Dalgarno (2001)]{bal01} Balakrishnan, N. \& Dalgarno, A. 2001, 
Chem.\ Phys.\ Lett., 341, 652

\bibitem[Brown and Balint-Kurti]{bro00} Brown, A., and Balint-Kurti, G.~G.\ 2000, 
J.\ Chem.\ Phys.\, 113, 1870

\bibitem[Castillo et al. (1998)]{cas98}Castillo, J.~F., Hartke, B., Werner, H.--J., Aoiz, F.J., Banares, L., Martinez-Haya, B. 1998, J.\ Chem.\ Phys.\ 109, 7224

\bibitem[Caux et al.(2002)]{2002A&A...383L...9C} Caux, E., Ceccarelli, C., 
Pagani, L., Maret, S., Castets, A., \& Pardo, J.~R.\ 2002, \aap, 383, L9 

\bibitem[Curran et al.(2004)]{2004MNRAS.352..563C} Curran, S.~J., Murphy, 
M.~T., Pihlstr{\" o}m, Y.~M., Webb, J.~K., Bolatto, A.~D., \& Bower, G.~C.\ 
2004, \mnras, 352, 563 

\bibitem[Dalgarno, Yan, \& Liu(1999)]{1999ApJS..125..237D} Dalgarno, A., 
Yan, M., \& Liu, W.\ 1999, \apjs, 125, 237 

\bibitem[Draine(1978)]{draine78} Draine, B.~T.\ 1978, ApJS, 36, 595 

\bibitem[Federman, Sheffer, Lambert, \& Smith(2005)]{2005ApJ} 
Federman, S.~R., Sheffer, Y., Lambert, D.~L., \& Smith, V.~V.\ 2005, 
ApJ, 619, 884 

\bibitem[Gredel et al. (1989)]{gre89} Gredel,  R., Lepp, S., Dalgarno,  A., \& Herbst, E.\ 1989, ApJ, 347, 289

\bibitem[Habing(1968)]{habing68}Habing, H. J. 1968, Bull.\ Astron.\ Inst.\
    Netherlands, 19 ,421

\bibitem[Hitchcock et al. (1984)]{hit84} Hitchcock, A.~P., Williams, G.~R.~J., Brion, C.E., \& Langhoff, P.~W. 1984, Chem. Phys. 88, 65

\bibitem[H]{H} Hollenbach, D.~J., et al.\ 2004, in preparation

\bibitem[J]{J} Jones, A.~P., \& Williams, D.~A. 1984, \mnras, 209, 955

\bibitem[Kaufman et al.(1999)Kaufman, Wolfire, Hollenbach, \& 
   Luhman]{kauf99} Kaufman, M.~J., Wolfire, M.~G., 
   Hollenbach, D.~J., \& Luhman, M.~L.\ 1999, \apj, 527, 795 (K99)

\bibitem[K]{K} Kaufman, M.~J., et al.\ 2005, in preparation

\bibitem[Le Bourlot et al.(1993)Le Bourlot, Pineau Des Forets, Roueff, \& 
Flower]{lebour93} Le Bourlot, J., Pineau Des Forets, G., 
Roueff, E., \& Flower, D.~R.\ 1993, \aap, 267, 233 

\bibitem[Ledoux, Petitjean, \& Srianand(2003)]{2003MNRAS.346..209L} Ledoux, 
C., Petitjean, P., \& Srianand, R.\ 2003, \mnras, 346, 209 

\bibitem[Le Petit, Roueff, \& Le Bourlot(2002)]{lepetit02} Le 
Petit, F., Roueff, E., \& Le Bourlot, J.\ 2002, \aap, 390, 369 

\bibitem[Le Teuff et al. (2000)]{let00} Le Teuff,  Y.~H., Millar,  T.~J., \& Markwick,  A.~J.\ 
2000, A\&AS, 146, 157

\bibitem[Li \& Draine(2001)]{2001ApJ...554..778L} Li, A.~\& Draine, B.~T.\ 
2001, \apj, 554, 778 

\bibitem[Mathis, Mezger, \& Panagia(1983)]{mathis83} Mathis, 
J.~S., Mezger, P.~G., \& Panagia, N.\ 1983, \aap, 128, 212 

\bibitem[McCall et al.(2003)]{mccall03} McCall, B.~J., et al.\ 
2003, \nat, 422, 500 

\bibitem[Mezger, Mathis, \& Panagia(1982)]{mezger82} Mezger, 
P.~G., Mathis, J.~S., \& Panagia, N.\ 1982, \aap, 105, 372 

\bibitem[Nee et al. (1985)]{nee85} Nee, J.B., Suto, M., \& Lee, L.C.\ 1985, J.\ Phys.\ B, 18, 293

\bibitem[Neufeld, Zmuidzinas, Schilke, \& 
Phillips(1997)]{1997ApJ...488L.141N} Neufeld, D.~A., Zmuidzinas, J., 
Schilke, P., \& Phillips, T.~G.\ 1997, \apjl, 488, L141 (NZSP)


\bibitem[Pardo et al.(2001)]{pardo01} Pardo, J.~R., Cernicharo, J., \& Serabyn, E.\ 2001, IEEE Trans.\ Ant.\ and Prop., vol.\ 49, nr.\ 12

\bibitem[Parravano, Hollenbach, \& McKee(2003)]{par03} 
Parravano, A., Hollenbach, D.~J., \& McKee, C.~F.\ 2003, \apj, 584, 797 

\bibitem[Pequignot(1996)]{peq96}P\'equignot, D.\ 1990, \aap, 231, 499 (Erratum: 313, 1026) 

\bibitem[P]{P}  Peterson, K.~A., Woods, R.~C., Rosmus, P., Werner, H.-J.\ 1990,
J.\ Phys.\ Chem., 93, 1889

\bibitem[Prasad & Tarafdar (1983)]{pra83} Prasad, S.~S., \& Tarafdar, S.~P. 1983, ApJ, 267, 603

\bibitem[Reese et al. (2005)]{ree05} Reese, C., Stoecklin, T., Voronin, A., and Rayez, R.~C.\ 2005, \aap, 430, 1139

\bibitem[Rowe et al. (1985)]{row85} Rowe, B.~R., Marquette, J.~B., Dupeyrat, G., \& Ferguson, E.~E.\ 1985, Chem.\ Phys.\ Lett.\, 113, 403 

\bibitem[Savage, Drake, Budich, \& Bohlin(1977)]{1977ApJ...216..291S} 
Savage, B.~D., Drake, J.~F., Budich, W., \& Bohlin, R.~C.\ 1977, \apj, 216, 
291 

\bibitem[Scott et al. (1997)]{Sco97} Scott, G.~B.~I., Fairley, D.~A., Freeman, C.~G., \& McEwan, M.~J. 1997, Chem.\ Phys.\ Lett., 269, 88 

\bibitem[Snow \& York(1981)]{1981ApJ...247L..39S} Snow, T.~P.~\& York, 
D.~G.\ 1981, \apjl, 247, L39 

\bibitem[Stark & Werner (1996)]{sta96} Stark, K., \& Werner, H.--J. 1996, J.\ Chem.\ Phys.\ 1996, 104, 6515 

\bibitem[Sternberg et al. 1987]{ste87} Sternberg, A., Dalgarno, A., \& Lepp, S.\ 1987, ApJ, 320, 676

\bibitem[Stevens et al. (1989)]{ste89} Stevens, P.S., Brune, W.H., \& Anderson, J.G. 1989, J.\ Phys.\ Chem., 93, 4068

\bibitem[T]{T} Tashiro, L.~M., Ubachs, W., \& Zare, R.~N.\ 1989, J.\ Mol.\ Spectr., 138, 89  

\bibitem[Tielens \& Hollenbach(1985)]{th85} Tielens, 
A.~G.~G.~M.~\& Hollenbach, D.\ 1985, \apj, 291, 722 

\bibitem[Troe (1985)]{tro85} Troe, J.\ 1985, Chem.\ Phys.\ Lett., 122, 425

\bibitem[Troe (1987)]{tro87} Troe, J.\ 1987, J.\ Chem.\ Phys., 87, 2773

\bibitem[Troe (1996)]{tro96} Troe, J.\ 1996, J.\ Chem.\ Phys., 105, 6249

\bibitem[Weingartner \& Draine(2001)]{wd01} Weingartner, 
J.~C.~\& Draine, B.~T.\ 2001, \apjs, 134, 263 

\bibitem[Wiklind \& Combes(1999)]{1999hrrl.conf..202W} Wiklind, T.~\& 
Combes, F.\ 1999, ASP Conf.~Ser.~156: Highly Redshifted Radio Lines, 202 

\bibitem[Wolfire, Tielens, \& Hollenbach(1990)]{wolfire90} 
Wolfire, M.~G., Tielens, A.~G.~G.~M., \& Hollenbach, D.\ 1990, \apj, 358, 116

\bibitem[Wolfire, McKee, Hollenbach, \& Tielens(2003)]{wolfire03} 
Wolfire, M.~G., McKee, C.~F., Hollenbach, D., \& Tielens, A.~G.~G.~M.\ 
2003, \apj, 587, 278 


\bibitem[Zhu et al. 2002]{Zhu02} Zhu, C., Krems, R., Dalgarno, A., \& Balakrishnan, N.\ 2002, ApJ, 577, 795

\end{thebibliography}
\end{document}